\documentclass[useAMS,usenatbib]{mn2e}

\usepackage{color}
\usepackage{braket}
\usepackage{amsmath}
\usepackage{graphicx}
\usepackage{hyperref}
\usepackage{amssymb}

\newcommand{\be}{\begin{equation}}
\newcommand{\ee}{\end{equation}}
\newcommand{\beq}{\begin{eqnarray}}
\newcommand{\eeq}{\end{eqnarray}}
\newcommand{\f}{f_L}
\newcommand{\s}{\sigma_{\f}}
\newcommand{\ff}{\tilde{f}_{\tilde{L}}}
\newcommand{\model}{f^*(x) = Ax + B\left[ \log (1 + x) \right]^W + C}
\newcommand{\tp}{\theta, \phi}
\newcommand{\lk}{\lambda, \kappa}
\newcommand{\Nb}{N_{\rm bound}}
\newcommand{\Nf}{N_{\rm free}}
\newcommand{\nNb}{n_{\rm bound}}
\newcommand{\Eeff}{E_{\rm eff}}
\newcommand{\Rws}{R_{\rm ws}}
\newcommand{\Rrange}{R_{\mathrm{range}}}

\title[Mesoscopic pinning forces in NS crusts]{Mesoscopic pinning forces in neutron star crusts}
\author[Seveso et al.]{S.~Seveso$^{1,2}$, P.~M.~Pizzochero$^{1,2}$\thanks{E-mail:pierre.pizzochero@mi.infn.it}, F.~Grill$^{1,2}$, B.~Haskell$^{3}$\\
$^{1}$Dipartimento di Fisica, Universit\`a degli Studi di Milano, Via Celoria 16, 20133 Milano, Italy\\
$^{2}$Istituto Nazionale di Fisica Nucleare, sezione di Milano, Via Celoria 16, 20133 Milano, Italy\\
$^{3}$School of Physics, The University of Melbourne, Parkville, VIC 3010, Australia}

\begin{document}

\pagerange{\pageref{firstpage}--\pageref{lastpage}} \pubyear{2015}

\maketitle

\label{firstpage}

\begin{abstract}

The crust of a neutron star is thought to be comprised of a lattice of nuclei immersed in a sea of free electrons and neutrons. As the neutrons are superfluid their angular momentum is carried by an array of quantized vortices. These vortices can pin to the nuclear lattice and prevent the neutron superfluid from spinning down, allowing it to store angular momentum which can then be released catastrophically, giving rise to a pulsar glitch. A crucial ingredient for this model is the maximum pinning force that the lattice can exert on the vortices, as this allows us to estimate the angular momentum that can be exchanged during a glitch. In this paper we perform, for the first time, a  detailed and  quantitative calculation of the pinning force \emph{per unit length} acting on a vortex immersed in the crust and resulting from  the mesoscopic vortex-lattice interaction. We consider realistic vortex tensions, allow for displacement of the nuclei  and average over all possible orientation of the crystal with respect to the vortex. We find that, as expected, the mesoscopic pinning force becomes weaker for longer vortices and is generally much smaller than previous estimates, based on vortices aligned with the crystal. Nevertheless the forces we obtain still have maximum values of order $f_{\rm{pin}}\approx 10^{15}$ dyn/cm, which would still allow for enough angular momentum to be stored in the crust to explain large Vela glitches, if part of the star is decoupled during the event.

\end{abstract}


\section{Introduction}

The physics of the Neutrons Star (NS) crust plays a crucial role when attempting to model these objects. First of all the outer layers of the star provide a heat blanket that shields the hot interior and determines the observable thermal emission from the surface \citep{Gud}. The elastic properties of the crust are also crucial, as 'crust-quakes' have been invoked to explain a number of phenomena, such as magnetar flares \citep{TD}  and pulsar glitches \citep{Alparquake,quake2}. Furthermore the crust may sustain a large enough strain to build a 'mountain' that leads to detectable gravitational wave emission \citep{Bild98}. In this paper we focus on the relation between crustal physics and pulsar glitches.

Glitches are sudden increases in frequency (instantaneous to the accuracy of the data) of an otherwise smoothly spinning down radio pulsar. After the event there is, in many cases, an increase in the spin-down rate that relaxes exponentially back to the pre-glitch spin-down \citep{Espinoza1}. Soon after the first observations, the long timescales associated with this relaxation (up to months) were associated with the re-coupling of a loosely coupled superfluid component in the NS crust \citep{Baym}. Neutron superfluidity in NS interiors is, in fact, expected on a theoretical basis \citep{Migdal} as most of the star will be cold enough for neutrons to form Cooper pairs and behave as a superfluid condensate, that can flow with little or no viscosity relative to the 'normal' component of the crust. Furthermore, recent observations of the cooling of the young NS in the supernova remnant Cassiopea A are consistent with this picture \citep{casA1,casA2, CASAnew}.

A crucial aspect of superfluid dynamics is that the neutron condensate can only rotate by forming an array of quantized vortices, which determine an average rotation rate for the fluid. For the superfluid to spin-down it is necessary for vorticity to be expelled. If vortices are, however, strongly attracted to the ions in the crust (i.e. they are 'pinned') their motion is impeded and the superfluid cannot follow the spin-down of the crust, and stores angular momentum, releasing it catastrophically during a glitch \citep{Itoh}.

The nature of the trigger for vortex unpinning is still debated, with proposals ranging from vortex avalanches \citep{Alpardeplete,Warszawski3} to hydrodynamical instabilities \citep{hydro} or crust quakes  \citep{Rude,Rude2,Alparquake,quake2}.
Whatever the trigger mechanism, an important ingredient in this picture is the maximum pinning force that the crust can exert on a vortex, before hydrodynamical lift forces (the Magnus force) are able to free it. This quantity obviously determines the maximum amount of angular momentum that can be exchanged during a glitch. An understanding of how much angular momentum can be stored in different regions of the star would, in fact, allow detailed comparisons with observations of glitching pulsars and potentially constrain the equation of state of dense matter \citep{crust, chamelcrust, chuck}.

Early theoretical work focused on the microscopic interaction between a vortex and a single pinning site \citep{Alpar77, EB}. The pinning force \emph{per unit length} of a vortex depends, however, on the mesoscopic interaction between the vortex and many pinning sites, and thus on the rigidity of the vortex, on its radius (represented by the superfluid coherence length $\xi$) and on the lattice spacing. This naturally leads to the possibility of different pinning regimes in different regions of the crust. 
\citet{Alpar84a,Alpar84b} interpreted the slow post-glitch recovery of the Vela pulsar in terms of vortex 'creep', i.e. thermally activated motion of pinned vortices, and distinguished between three regimes: strong, weak and super weak pinning.
The different regimes depend on the interplay between the quantities mentioned earlier: in strong pinning the coherence length $\xi$ of a vortex is smaller than the lattice spacing, and the interaction is strong enough to displace nuclei; while in the weak pinning regime this is not the case. Superweak pinning, on the other hand, comes about when the coherence length $\xi$ is greater than the lattice spacing and a vortex can encompass several nuclei. In this case there is little change in energy as the vortex moves and thus no preferred configuration for pinning. The pinning force is expected to be weak and, in the limit of infinitely long vortices all configurations are equal and there would be no pinning \citet{jonesnopin}.
Fits to the post-glitch relaxation of the Vela pulsar, within the vortex creep framework \citep{Alpar84a}, were used to set observational constraints on some of these parameters, leading to the conclusion that only weak and super weak pinning are likely to be at work in a neutron star crust \citep{Alpar84b}. 
The theoretical calculations of the mesoscopic pinning force relied, however, on estimates in the weak pinning case for the very particular configuration of vortices \emph{aligned} with the crystal axis. Although very little is known about the defect structure of the crust, one does not in general expect the crystal lattice to be oriented in the same direction over the whole length of a vortex (note also that a vortex will have cylindrical symmetry set by the rotation axis, while the only preferred direction for the crystal will be set by gravity and pressure which have spherical symmetry, slightly modified by rotation).
More recently \citet{link2009} has performed simulations of motion of a vortex in a three-dimensional random potential, and found that the rigidity of the vortex does, indeed, play a fundamental role in setting the maximum superfluid flow above which vortices cannot remain pinned. \citet{linkcutler} and \citet{link} also estimated the pinning force per unit length of a rigid vortex in a random lattice using a variational approach, also including phenomenologically the effect of strong entrainment in the crust \citep{link14}.

In this paper we perform a realistic calculation of the mesoscopic pinning force, that is the force per unit length acting on straight vortices in the neutron star crust. We consider for the first time a micro physically motivated model for the crust and investigate the density dependence of the pinning force. We average over all possible vortex-crystal orientations and show that, although the force is considerably weaker than previous estimates based on particular configurations, it could still be strong enough to account for angular momentum transfer in large pulsar glitches.

\section{Lattice properties}
\label{sec:latticeprop}

The crust of a NS is thought to form a crystal in which completely ionized neutron-rich nuclei form a {\em body centered cubic} (BCC) lattice, immersed in a sea of electrons and free neutrons. In this configuration each nucleus is at the centre of a cubic cell of side $s=2\Rws$ with nuclei at each vertex. The separation between the ions (i.e. the potential pinning sites) thus depends on $\Rws$, the radius of the Wigner-Seitz cell, which is a function of the density $\rho$. In our calculation we use the classic results from \citet{Negele}  where the crust is divided in five zone, each one characterized by a specific value of $\Rws$ and $R_N$, which is the radius of the nucleus that occupies a single site of the lattice. Table \ref{tab:negele} summarizes these results, together with the nuclear composition of the Wigner-Seitz cells.
\begin{table*}
\centering
\setlength{\tabcolsep}{14.5pt}
\caption{Fiducial values of the quantities used in our calculations. These values are taken from \citet{Negele}: the NS crust is divided in five zone and here we give the baryon density $\rho$, the Wigner-Seitz cell radius ($\Rws$), the element corresponding to the cell nuclear composition, the nuclear radius ($R_N$), the superfluid coherence length ($\xi$), which represents the vortex radius, and the pinning energy per site ($E_p$). The last two quantities are taken from the results of \citet{DP04,DP06}}
\label{tab:negele}
\begin{tabular}{rrrrrrlcrr}
	\hline
	\# & $\rho$ [g cm$^{-3}$] &                      Element & $\Rws$ [fm] & $R_N$ [fm] & \multicolumn{2}{c}{$\xi$ [fm]} &  & \multicolumn{2}{c}{$E_p$ [MeV]} \\ \cline{6-7}\cline{9-10}
	   &                      &                              &             &            & $\beta = 1$ & $\beta = 3$           &  & $\beta = 1$ &          $\beta = 3$ \\ \hline
	 1 & $1.5 \times 10^{12}$ &   $^{320}_{\phantom{0}40}$Zr &        44.0 &        6.0 &       6.7 & 20.0                &  &      2.63 &               0.21 \\[4pt]
	 2 & $9.6 \times 10^{12}$ & $^{1100}_{\phantom{00}50}$Sn &        35.5 &        6.7 &       4.4 & 13.0                &  &      1.55 &               0.29 \\[4pt]
	 3 & $3.4 \times 10^{13}$ & $^{1800}_{\phantom{00}50}$Sn &        27.0 &        7.3 &       5.2 & 15.4                &  &     -5.21 &              -2.74 \\[4pt]
	 4 & $7.8 \times 10^{13}$ & $^{1500}_{\phantom{00}40}$Zr &        19.4 &        6.7 &      11.3 & 33.5                &  &     -5.06 &              -0.72 \\[4pt]
	 5 & $1.3 \times 10^{14}$ &   $^{982}_{\phantom{0}32}$Ge &        13.8 &        5.2 &      38.8 & 116.4               &  &     -0.35 &              -0.02 \\[4pt] \hline
\end{tabular}
\end{table*}

Note that there is still significant uncertainty on the exact composition and structure of the crust \citep{steiner, chuck} and not only electrons, but also free neutrons, may partially screen the Coulomb interaction between the nuclear clusters, leading to different, and more inhomogeneous, configurations than a BCC lattice \citep{Pethick}. Nevertheless the procedure we describe below can easily be adapted to different configurations.

To calculate the mesoscopic pinning force we need to identify the configurations  in which the vortex is most strongly pinned to the lattice and the configurations in which it is 'free' (note that a vortex is never truly free, as it will always intersect pinning centres. The `free' configuration simply represents the average energy configuration  between locations of maximum pinning, as discussed in the next section). Once this has been done the {\it maximum} pinning force $F_p$ simply follows from:
\be
F_p=\frac{E_{\mathrm{free}}-E_{\mathrm{pin}}}{\Delta r}
\ee
where $E_{\mathrm{pin}}$ is the energy of the most strongly pinned configuration and $E_{\mathrm{free}}$ the energy of the free configuration. The average distance the vortex has to move between the configurations is $\Delta r$. 

The energy of a particular vortex configuration will depend on the number of ions that it is able to pin to. Intuitively, the more sites it can pin to, the greater the energy gain, the stronger the pinning. In order to perform the calculation it is thus necessary to consider the pinning energy {\it per pinning site} $E_p$, i.e. the amount by which the energy of the system is changed when a single nucleus is inside the vortex. This quantity depends on the competition between the kinetic energy and the condensation energy of the superfluid, which is strongly density dependent and will thus change if a dense nucleus is introduced in the vortex. In this work we use the results of \citet{DP03,DP04,DP06}, who calculate $E_p$ consistently in the local density approximation. The values of $E_p$ for different densities are given in the last columns of table \ref{tab:negele}. Note that in some regions $E_p$ is positive, i.e. it costs energy to introduce a nucleus in a vortex. In these regions the vortex-nucleus interaction is repulsive and one has 'interstitial' pinning (IP), in which the favored vortex configurations are in-between nuclei. We refer to the case in which the interaction between nuclei and vortices is attractive as 'nuclear' pinning (NP). We shall see in the following that the effect of attraction or repulsion does not strongly influence the calculation of the mesoscopic pinning force. The parameter $\beta$ refers to the suppression factor for the neutron pairing gap used in the calculations: $\Delta = \frac{\Delta_0}{\beta}$, where $\Delta_0$ is the pairing gap of the superfluid obtained by using the bare interaction (i.e. not accounting for in-medium corrections). This factor is related to the polarization effects of matter on the nuclear interaction. The case $\beta = 1$ describes the non--polarized interaction, while the case $\beta = 3$ describes the one in which the effect of the polarization is maximum. When $\beta = 1$ the mean pairing gap has a maximum of about $3$ MeV, which corresponds to the strong pairing scenario, while when $\beta = 3$ the mean pairing gap has a maximum of about $1$ MeV, as usually assumed in the weak pairing scenario. Realistic Montecarlo simulations of neutron matter \citep{Gandolfi2008} indicate a reduction of the pairing gap consistent with the choice $\beta = 3$.

The total energy of the interaction between a given vortex portion and the lattice is calculated summing the contribution of each nucleus that can be captured by the pinning force. Naively this could be done by considering the vortex as a cylinder of radius $\xi$ and counting how many nuclei are contained within it (we will discuss how to count nuclei at the boundary in the following). This approach can be improved to take into account the possible deformation of the nuclear lattice. The lattice has elastic properties, so it is possible for nuclei to be displaced from their equilibrium position under the action of the pinning force. The resulting energy per site can be expressed as
\begin{equation}
\label{eq:p_energy}
E(r) = E_p + E_l(r)
\end{equation}
where $r$ is the distance of the  vortex axis from the equilibrium position of the considered nucleus. In this approach, the pinning energy per site $E_p$ is corrected by the factor $E_l(r)$ that encodes the change in electrostatic energy due to the displacement of the nucleus. We will then define the capture radius $r_c$ as the radius within which it is energetically favorable for the nuclei to be displaced: this will be the radius of the vortex to be used in the counting procedure. Let us now estimate $r_c$ for both nuclear and interstitial pinning.

\subsection{Nuclear pinning}

\begin{figure}
	\centering
    \def\svgwidth{4cm}
    \input{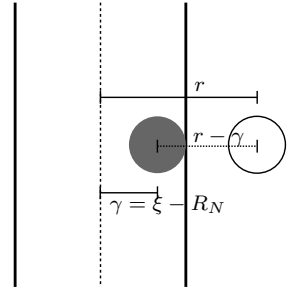}
	\caption{Representation of a nucleus displacement (NP case). The empty and full circles represent respectively the starting and final position of the nucleus. The dashed line represents the displacement $\delta(r)$.}
	\label{fig:np_scheme}
\end{figure}
In the nuclear pinning regime ($E_p < 0$) we define a pinning region assuming that a nucleus contributes to the total interaction by a factor $E_p$ if it is completely inside the vortex: in other words its distance from the vortex axis must be  less than $\gamma = \xi - R_N$ (figure \ref{fig:np_scheme}), with $\gamma =0$ if $\xi < R_N$. If a site is at a distance $r>\gamma$ from the vortex axis, the nucleus must be dragged by a distance $\delta(r) = r - \gamma$.
The electrostatic energy is calculated in a standard way using  Gauss theorem together with the Wigner-Seitz approximation, which divides the lattice in independent spherical cells
of radius $\Rws$ each with an ion in the center surrounded by the electron and neutron gas:
\begin{equation}
E_l(r) = \frac{Z^2 e^2}{2 \Rws^3} \delta^2(r)
\end{equation}
where $e$ is the elementary charge and $Z$ is the number of protons and electrons in the cell. Of course, a nucleus whose equilibrium position is already inside the pinning region does not need to be dragged, so its energy contribution has no electrostatic term ($E(r) = E_p$ if $r<\gamma$). We can now define the maximum drag distance $r_0$ as the value of $\delta(r)$ for which the effective pinning interaction of equation (\ref{eq:p_energy}) becomes zero: 
\begin{equation}
r_0 = \sqrt{-\frac{2 E_p \Rws^3}{Z^2 e^2}}.
\end{equation}
>From these consideration, it follows that the final {\em capture radius} that must be used in our calculation will be
\begin{equation}
r_c = \gamma  + r_0 = \xi - R_N + r_0
\end{equation}

The total energy of the interaction between the considered vortex portion (of length $L$) and the lattice is calculated summing the contribution of each nucleus that can be captured by the pinning force. 
This energy is calculated through an integral over a uniform distribution of nuclei, that is valid when the number of nuclei which are taken into account becomes very large, so for $L \gg \Rws$. Given $N$ the number of pinning sites that fall inside a cylinder of radius $r_c$ and length $L$, the superficial density will be $n_N = \frac{N}{\pi r_c^2}$. Then the total energy is calculated as 
\begin{eqnarray}
E &=& \int_0^\gamma E_p n_N\, 2\pi r\, dr + \int_\gamma^{\gamma+r_0} \left(E_p + E_l(r)\right) n_N\, 2\pi r\, dr \nonumber \\
 &=& \frac{N E_p}{(\gamma + r_0)^2}\left(\gamma^2 + \frac{4}{3}\gamma r_0 + \frac{1}{2}r_0^2 \right)
\end{eqnarray}
>From this equation we can immediately evaluate the effective interaction energy \emph{per site} $\Eeff$, defined by $E = N \Eeff$:
\begin{equation}
\Eeff = \frac{E_p}{(\gamma + r_0)^2}\left(\gamma^2 + \frac{4}{3}\gamma r_0 + \frac{1}{2}r_0^2 \right)
\end{equation}
In table  \ref{tab:latticeprop} we give the values of the above quantities, which have been calculated using the fiducial inner crust and superfluid properties of table \ref{tab:negele}.

\subsection{Interstitial pinning}
\begin{figure}
	\centering
    \def\svgwidth{4cm}
    \input{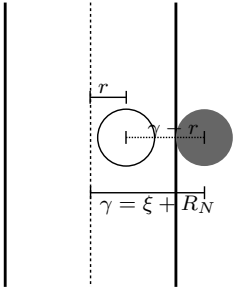}
	\caption{Representation of a nucleus displacement (IP case). The empty and full circles represent respectively the starting and final position of the nucleus. The dashed line represents the displacement $\delta(r)$. }
	\label{fig:ip_scheme}
\end{figure}
The evaluation of $r_c$ and $\Eeff$ in the  interstitial pinning regime  ($E_p > 0$) follows the same steps of the previous section, but taking into account the fact that in this case the interaction is repulsive and thence a nucleus that lies in the vortex core must be expelled instead of dragged into it in order to lower the energy. We define a nucleus as expelled if it is completely outside the vortex, that is if its distance from the vortex axis is larger than $\gamma = \xi + R_N$  (figure \ref{fig:ip_scheme}); a nucleus which is expelled does not contribute to the pinning energy. The drag distance now is $\delta(r) = \gamma - r$ and the maximum value for this quantity, $r_0$, is given by the energy balance $E_p = E_l(\delta = r_0)$. This encodes the idea that the nuclear displacement is favorable until the energy of the dragged nucleus configuration is lower than the energy of the configuration where the nucleus is still in its equilibrium position in the lattice:
\begin{equation}
r_0 = \sqrt{\frac{2 E_p \Rws^3}{Z^2 e^2}}.
\end{equation}
The capture radius that must be used in the counting procedure in this case is  equal to $\gamma$ because the nuclei that contribute to the pinning energy are only those that lie in the pinning region
\begin{equation}
r_c = \gamma = \xi + R_N
\end{equation}
Now, if $r_0 < \gamma$ the total energy is calculated as
\begin{equation}
E = \int_0^{\gamma-r_0} E_p n_N\, 2\pi r\, dr + \int_{\gamma-r_0}^\gamma E_l(r) n_N\, 2\pi r\, dr
\end{equation}
where the second term of the integral contains only the electrostatic contribution because the nuclei in that region have been expelled. If instead $r_0 > \gamma$ all the nuclei that contribute to the pinning energy are dragged outside the vortex: in this case we have
\begin{equation}
E = \int_0^{\gamma} E_l(r) n_N\, 2\pi r\, dr
\end{equation} 
Solving these integrals and defining again $E = N \Eeff$ we obtain the effective pinning energy \emph{per site} (see table \ref{tab:latticeprop} for numerical results):
\begin{equation}
\Eeff = 
\left\lbrace
\begin{array}{ll}
E_p\frac{1}{\gamma^2}\left(\gamma^2 - \frac{4}{3}\gamma r_0 + \frac{1}{2}r_0^2 \right) & r_0 \le \gamma \\
E_p\frac{\gamma^2}{6r_0^2} & r_0 > \gamma
\end{array}
\right.
\end{equation}

\begin{table*}
\centering
\caption{Lattice properties for the five zones of table \ref{tab:negele}. The values in table \ref{tab:negele} are used here to calculate the capture radius $r_c$ (in units of $\Rws$) and the effective pinning energy per site $\Eeff$  as explained in section \ref{sec:latticeprop}}
\label{tab:latticeprop}
\begin{tabular}{lcrrrrrrrrrrr}
\hline
& & \multicolumn{5}{c}{$\beta = 1$} & & \multicolumn{5}{c}{$\beta = 3$} \\
\cline{3-7} \cline{9-13}\\[-4pt]
\# & IP/NP & 
$\gamma$ [fm] & $r_0$ [fm] & $r_c$ [$\Rws$] & $E_p$ [MeV] & $\Eeff$ [MeV] & &
$\gamma$ [fm] & $r_0$ [fm] & $r_c$ [$\Rws$] & $E_p$ [MeV] & $\Eeff$ [MeV] \\
\hline
1 & IP & 12.7 & 14.0 & 0.289 &  2.63 &  0.36 &&  26.0 & 3.9 & 0.591 &  0.21 &  0.17 \\
2 & IP & 11.1 &  6.2 & 0.313 &  1.55 &  0.64 &&  19.7 & 2.7 & 0.555 &  0.29 &  0.24 \\
3 & NP &  0.0 &  7.6 & 0.204 & -5.21 & -2.60 &&   8.1 & 5.5 & 0.504 & -2.74 & -2.08 \\
4 & NP &  4.6 &  5.7 & 0.531 & -5.06 & -3.46 &&  26.8 & 2.1 & 1.490 & -0.72 & -0.69 \\
5 & NP & 33.6 &  1.1 & 2.514 & -0.35 & -0.34 && 111.2 & 0.3 & 8.080 & -0.02 & -0.02 \\
\hline
\end{tabular}
\end{table*}

\subsection{Vortex length}
\label{sec:vlen}
\begin{figure}
	\centering
    \def\svgwidth{8cm}
    \input{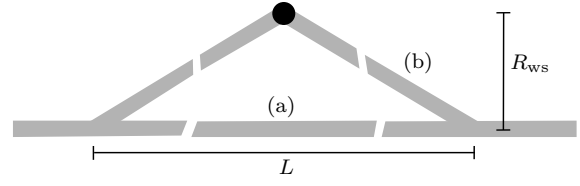}
	\caption{Representation of the vortex deformation. We sketch a rigid
vortex (a) and a bent vortex (b). $L$ is the vortex length and $\Rws$ is the Wigner-Seitz radius. Note that a realistic vortex configuration will not have any kinks, and these appear here simply for ease of plotting.}
	\label{fig:length}
\end{figure}
The length-scale over which a vortex can be considered straight corresponds the length $L$ of the cylinder on which we perform  the counting procedure described. We can estimate the order of magnitude of $L$ with a simple argument based on energy considerations (we develop the argument in the NP regime, but the same result obtains in the IP regime).  Assuming that the vortex, under tension $T$ (self-energy per unit length), will bend under the influence of the pinning force, we can equate the energy of two limiting configurations: the straight (infinitely rigid) vortex (figure \ref{fig:length}a) and the vortex that has bent in order to pin to an additional nucleus at a typical distance of order $\Rws$ (figure \ref{fig:length}b):
\begin{equation}
T L = E_p + T(L + \Delta L) 
\end{equation}
The difference $\Delta L$ of the vortex length in the two configuration is obviously
\begin{equation}
\Delta L = 2\sqrt{\left(\frac{L}{2} \right)^2 + \Rws^2} - L \approx \frac{2\Rws^2}{L}
\end{equation}
where we have expanded the expression following the realistic assumption that $\Rws \ll L$. Finally we have 
\begin{equation}
\frac{L}{\Rws} = \frac{2T\Rws}{|E_p|} \sim 10^3 \label{eq:length}
\end{equation}
where the standard neutron star values have been used: $T \sim 20 \mbox{ MeV fm}^{-1}$ \citep{Jones1990}, $\Rws \sim 30 \mbox{ fm}$ and $|E_p| \sim 1 \mbox{ MeV}$. We will thus study the dependence of our results on variations of the parameter $L$ around the estimate in equation (\ref{eq:length}). Note that the ability of a vortex to bend and adapt to a pinned configuration plays an important role in determining the maximum of the pinning force, as also found by \citet{link2009}. In particular we shall see in section \ref{sec:length} that our estimate of  $T \sim 20 \mbox{ MeV fm}^{-1}$ is appropriate for high density regions at the base of the crust, but that the tension can be much lower at lower densities, leading to vortices being much less rigid and higher values of the pinning force.

Let us point out that this is the length-scale over which a vortex would need to unpin to move to a different configuration, and agrees with the estimate of \citet{link14} and  with the numerical simulations of \citet{HS01}. 
In Appendix \ref{append} we also present a hydrodynamical calculation of the length-scale over which a rigid vortex can unpin, which agrees with the estimate for $L$ in equation (\ref{eq:length}). It should not be confused with the average length-scale between pinning bonds, $l_p$, defined by \citet{linkcutler} and \citet{link} which has a different scaling with the parameters. We shall see in the following that our functional form for the pinning force per unit length scales with the tension in the same way as that of \citet{link} and will interpret $l_p$ accordingly.

We point out, however, that $l_p$ cannot represent  the \emph{actual} distance between successive nuclear pinning sites as assumed in \citet{linkcutler} and \citet{link}. Indeed, these author define the rigidity length $l_p$ by the additional constraint that $\pi u^2 l_p n_N=1$, with $n_N \approx 1/R_{ws}^3$, to determine  the vortex lateral deviation $u$: namely the vortex length must be such that bending defines a volume that contains only one nucleus. In order for this to be consistent the deviation $u$ must be larger than the vortex core $\xi$ for point--like nuclei, as also explicitly noted in \citet{linkcutler} (for real nuclei, one should require $u > \xi + R_N$). With standard parameters for the inner crust, however, this consistency condition is never satisfied, making the approach of \citet{linkcutler} and \citet{link} physically incorrect. We will come back to this at the end of section \ref{sec:random}.

We finally notice that the choice $u = \Rws$  made in the derivation of equation (\ref{eq:length}) is physically reasonable for a general, order-of-magnitude definition of vortex rigidity since $\Rws$ is the natural length-scale of the system, also when pinning is concerned. For example, the pinning energies per site $E_p$ calculated so far and used to estimate the mesoscopic pinning forces correspond to the difference in energy between two configurations (nuclear and interstitial), where the vortex has moved by a distance $\Rws$ \citep{DP03,DP04,DP06}. Moreover, since $2\Rws$ is the average distance between nuclei, displacing a segment of vortex by a distance of order $\Rws$ will likely intersect a new nucleus. Of course, alternative and more specific choices of the typical deviation $u$ could also be proposed as plausible: for instance, $u=\xi + R_N$ or $u=r_c$. From tables \ref{tab:negele} and \ref{tab:latticeprop}  we see that these choices would give deviations that, depending on the density and for the more realistic case $\beta =3$ of weak pinning, can be either larger or smaller than $\Rws$ by a factor of less than two\footnote{The deepest region, zone 5, has both $\xi  , r_c \gg \Rws$ so that the choices $u=\xi + R_N$ or $u=r_c$ would give very large $L$ and thence very small pinning forces, corresponding to the superweak pinning regime  of \citet{Alpar84a}. Also the choice $u=\Rws$, however, leads to very weak pinning at high densities once we consider the density dependence of the tension, as discussed in section \ref{sec:length} and shown in figure (\ref{fig:fitL}).}.
Because of this ambiguity in the definition of $u$, in the following we will show results associated to a range of values of $L$. However, rescaling the deviation as $u=\alpha \Rws$ implies rescaling  the rigidity length $L$ by a factor $\alpha^2$  while,  due to the weak dependence  $f_L \sim 1/\sqrt{L}$ shown in equation (\ref{eq:poisson}), the mesoscopic pinning force $f_L$ is rescaled by a factor $\alpha^{-1}$.  
Therefore, although the uncertainty in the choice of $u$ implies uncertainties  in the final pinning forces, we expect this error to be less than a factor of two so that the orders of magnitudes and relative strengths estimated for $f_L$ will still be quite reliable.

\section{Mesoscopic pinning force}

The calculation of the pinning force per unit length is done here by \emph{counting} the actual number of pinning sites intercepted by a randomly oriented vortex. 
\begin{figure}
	\centering
	\vspace{0.5cm}
	\includegraphics[width=8cm]{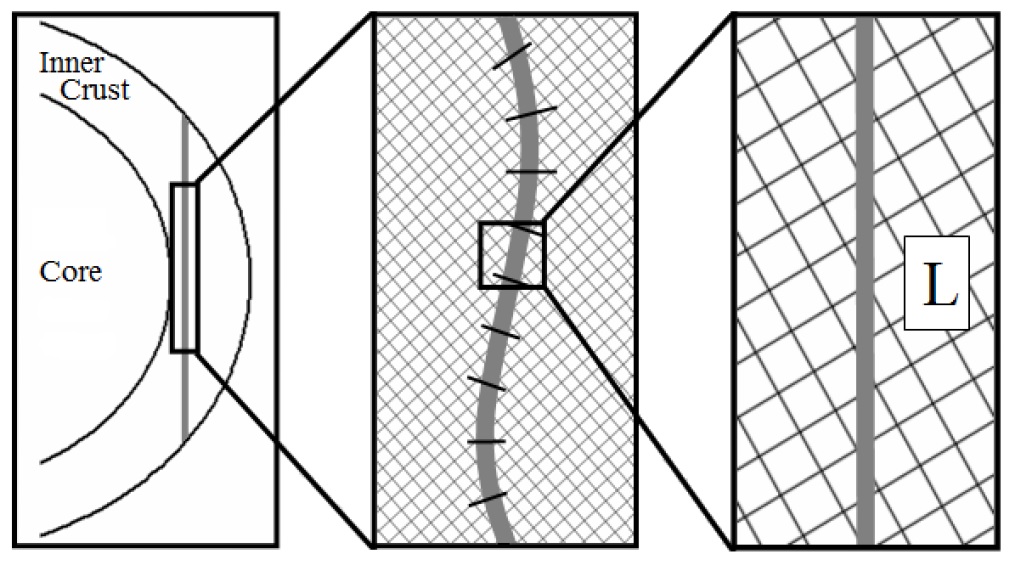}
	\caption{Representation of the vortex rigidity on different scales. $L$ is the maximum length of the unbent vortex as discussed in section \ref{sec:vlen}.}
	\label{fig:rigidity}
\end{figure}

We consider vortices parallel to the rotation axis and that thread the whole star. Due to the finite rigidity of the vortex we assume that it can be considered straight only on a characteristic length-scale $L$, as described in the previous section (figure \ref{fig:rigidity}). This idea, combined with the fact that the lattice is made up by macro-crystals with random direction \citep{Jones1990}, indicates that a macroscopic portion of vortex immersed in the crust experiences all possible orientations with respect to the lattice. The force per unit length should then be calculated as an average over all angular directions.
In following this procedure we neglect the effects of turbulence, which may arise in NS interiors \citep{Peralta05, Peralta06, Andersson07}, possibly due to modes of oscillations of the superfluid that may be unstable in the presence of pinning \citep{hydro,link}. In this case the vortex array is likely to form a complex tangle, that must, however, still be polarized due to the rotation of the star. Given that the problem of polarized turbulence is poorly understood (see \citet{Andersson07} for the description of a possible approach to this issue) we shall focus on a regular vortex array in this paper, and leave the complex problem of turbulence for future work.

\begin{figure*}
	\centering
	\includegraphics[width=14cm]{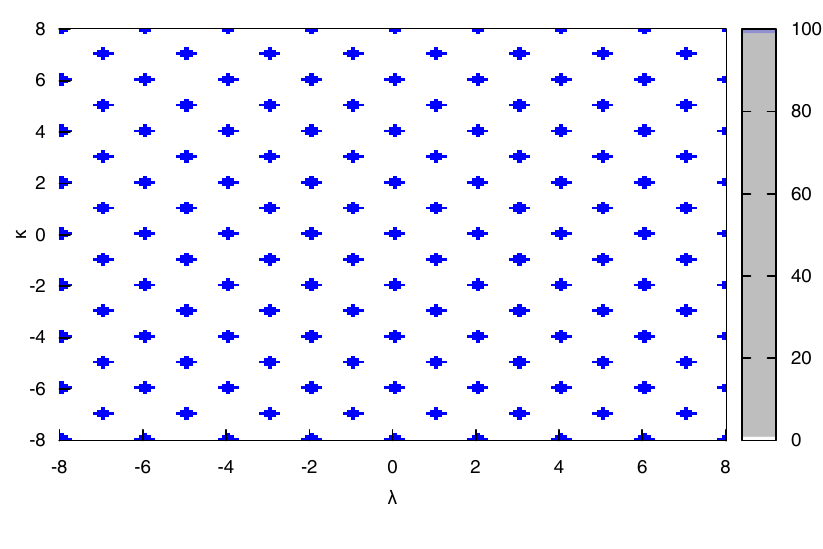}
	\caption{Number of captured pinning sites $N(\lk)$ when the vortex is \emph{aligned} with the lattice. The color codes are described in the sidebar. The axes $\lambda$ and $\kappa$ represent the translation of the vortex with respect to the initial position, and they are measured in $\Rws$ units. The grid step size is $dh = 0.1 \Rws$, the vortex is $L = 200 \Rws$ and the capture radius is $r_c = 0.204\Rws$ (region 3 with $\beta = 1$). Note that the simple geometry leads to several disjoint maxima that spread over several steps, given that for a small grid step $dh$ the energy of the configuration does not change until the vortex has been moved by one capture radius away from the aligned nuclei. }
	\label{fig:densityAl}
\end{figure*}

\begin{figure*}
	\centering
	\includegraphics[width=14cm]{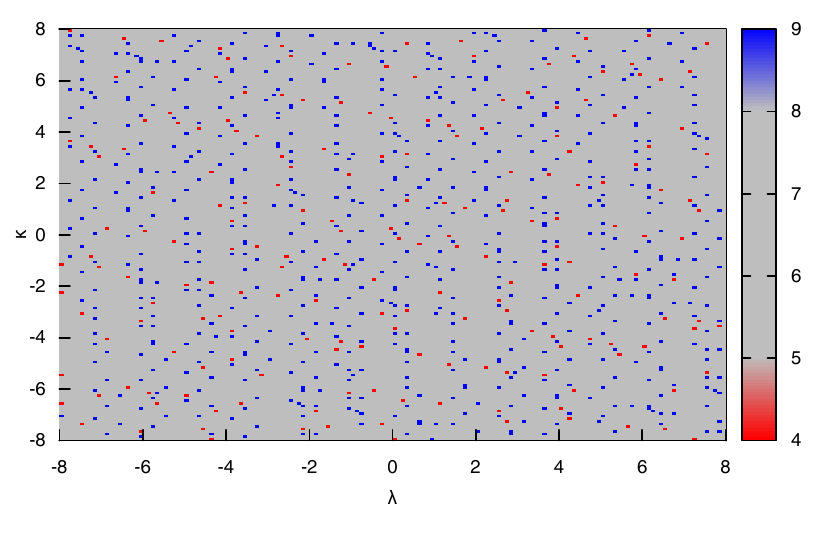}
	\caption{Number of captured pinning sites $N(\lk)$ when the vortex is \emph{non--aligned} with the lattice (we selected a random orientation). The color codes are described in the sidebar. For details on the parameters used to produce this plot see figure \ref{fig:densityAl}.  }
	\label{fig:densityNAl}
\end{figure*}

We consider an infinite BCC lattice with its symmetry axes oriented as $\hat{x}$, $\hat{y}$ and $\hat{z}$, and with a nucleus in $(0,0,0)$. A vortex is modeled as a cylinder of length $L$ and radius $r_c$ with its median point initially in the origin and the orientation is given by the angles $\theta$ and $\phi$ in spherical coordinates. For a given choice of $\theta$ and $\phi$, we evaluate the pinning force per unit length $f_L(\tp)$ by a counting procedure: from the initial position the vortex is moved parallel to itself, covering a square region of side $l$ in the plane perpendicular to the vortex axis, with steps of an amount $dh$. For each new position, identified by the displacement $(\lk)$, it is possible to count the number $N(\lk)$ of lattice nuclei that are within the capture radius of the vortex. In figures \ref{fig:densityAl} and \ref{fig:densityNAl},  we show two examples of a density plot where for each translation of the vortex $(\lk)$ we plot the number of captured pinning sites $N(\lk)$. The difference between the cases of vortex aligned with the lattice and vortex with arbitrary orientation is evident from the figures.

\begin{figure*}
	\centering
	\includegraphics[width=14cm]{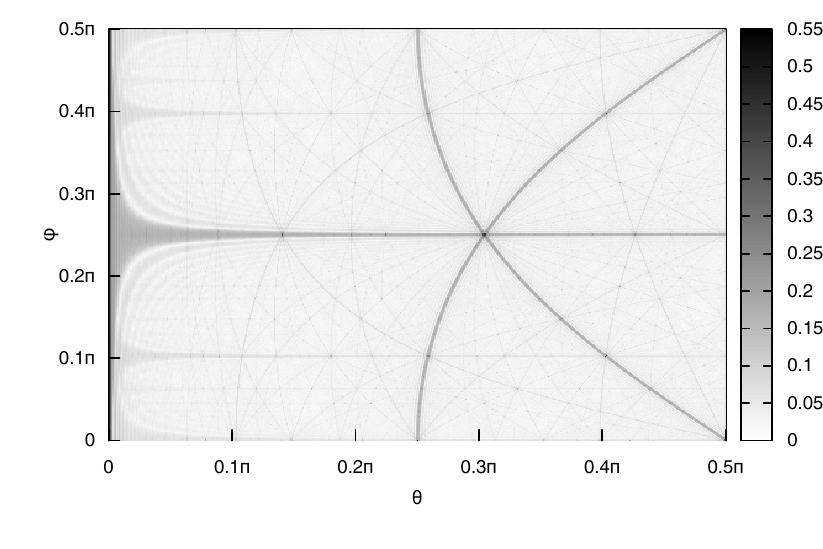}
	\caption{Difference between the number of pinning sites of the free and bound configurations as a function of the vortex orientation $(\tp)$. Here we plot  $|\Delta N(\tp)|/\tilde{L}$, where $\tilde{L}$ is the (adimensional) vortex length in units of  the Wigner-Seitz radius. The color codes are described in the sidebar. The figure has been obtained considering a vortex of length $L = 200 \Rws$ and capture radius $r_c = 0.204 \Rws$ (region 3 with $\beta = 1$).}
	\label{fig:density_theta_phy}
\end{figure*}

As discussed in the previous section, the number of captured pinning sites $N$ in a vortex-lattice configuration is directly related to the energy of the configuration by the expression $E = \Eeff N$ where $\Eeff$ is the effective contribution of every single interaction. As previously discussed, the interaction between the vortex and the nuclei can be attractive (NP) or repulsive (IP) in different regions of the crust. The calculation procedure presented here is valid for both cases, with the following distinction: in the NP regime, the {\em bound} configuration (state of minimum energy) is identified by the positions $(\lk)$ where the number of pinning site reach its maximum. This means that $\Nb(\tp) = \max(N(\lk))$. On the other hand, in the IP case, we must take the minimum: $\Nb(\tp) = \min(N(\lk))$.

This leads to the fact that, for both the NP and IP cases, the change in energy obtained by moving the vortex away from its bound configuration (unpinning energy) will be:
\begin{eqnarray}
\Delta E (\tp) &=& \Eeff \Delta N(\tp)  \nonumber \\
 &=& \Eeff \left( \Nf(\tp) - \Nb(\tp) \right),
\end{eqnarray}
where we take $\Nf$ as the average number of pinning sites counted in all visited displacements: $\Nf(\tp) = \braket{N(\lk)}$. Obviously,  we have $\Delta E (\tp) > 0$ for both the NP and IP cases, since it takes energy to remove the vortex from the  location where it is pinned.  We see that in any given zone (fixed $\Eeff$ and $r_c$) the unpinning energy depends on the vortex  orientation only through $\Delta N(\tp)=\Nf(\tp) - \Nb(\tp) $, the change in the number of captured nuclei between the two configurations. In figure \ref{fig:density_theta_phy} we plot the quantity $|\Delta N(\tp)|/\tilde{L}$ as a function of $(\tp)$, where $\tilde{L}=L/\Rws$ is the (adimensional) vortex length in units of $\Rws$; the plot corresponds to region 3 (NP regime), so that actually $\Delta N(\tp)<0$. Notice that the aligned configuration of figure \ref{fig:densityAl} would correspond to $\Delta N(0,0)/\tilde{L}=-0.5$, since $\Nf(0,0)=0$ and $\Nb(0,0)=L/2\Rws$ (the captured nuclei are a distance $s=2\Rws$ apart). It is evident from the figure that most orientations have $|\Delta N(\tp)|/\tilde{L}\ll 0.5$.

The force required to move the vortex away  from the bound configuration can be easily calculated using the following expression:
\begin{equation}
F(\tp) = \frac{\Delta E(\tp)}{D(\tp)}
\label{eq:force}
\end{equation}
where $D(\tp)$ identifies the average distance required to reach the free configuration from the pinned one. We estimate this quantity by counting in the density plot the number $\nNb$ of disjoint positions where $N(\lk) = \Nb$ (we sometimes omit the angular dependence for notational simplicity). In other words, $\nNb$ represents the number of distinct  extremal configurations (maxima in the NP regime, minima in the IP regime) found in the sampling square region. For a uniform distribution of these extremal points (square array of step $2D$) we would have $\nNb \pi D^2 \simeq l^2$, where $l$ is the side of the square region tested by parallel--transporting the vortex. We thus take as a reasonable definition for the average distance in the general case:   
\begin{equation}
\label{eq:distance}
D(\tp) = \frac{l}{\sqrt{\pi\nNb(\tp)}}
\end{equation}
Finally, the force per unit length is 
\begin{equation}
\f(\tp) =   \frac{F(\tp)}{L}
\end{equation}

For the procedure described above it is clearly necessary to unambiguously count $\nNb(\tp) $.
With the parallel--transport operation, we explore a portion of the plane that is perpendicular to the vortex axis. This region is a square region of side $l$ that is sampled with a grid of step $dh$. This means that we have to look for the position of maxima/minima  analyzing a set of points $(\lambda, \kappa)_{ij} = (-l/2 + i\, dh, -l/2 + j \, dh)$. If we merely count the number of  points for which $N_{ij} = N(\lambda, \kappa)_{ij}$ reaches its maximum/minimum value, this result would be strongly conditioned by the choice of the $dh$ parameter. In fact, for small values of $dh$, it is obvious that a single ``maximum/minimum position'' will be split over several points $(\lambda, \kappa)_{ij}$, altering the final result. 

One possible solution is to take into account only  {\em disjoint} maxima/minima: this means that two extremal points count as one if they are ``first--neighbors''. This approach requires a second--pass analysis over the values $N_{ij}$ to identify the clusters in the density map, and it allows us to correctly evaluate a configuration such as the one in figure \ref{fig:densityAl}, in which the alignment of the vortex with the symmetry axis of the crystal leads, for small grid steps, to several neighboring equivalent configurations. Without considering clustering, we would have counted $\nNb = 1745$  for this particular case ($dh = 0.1\Rws$). Counting only {\em disjoint} extremal points, instead, gives  the correct answer of $\nNb = 145$, and this result does not change if we explore the square region with a smaller step size.

In this work the method just described has been slightly generalized to treat extremal points that are topologically disjoint but ``very close'' and thence physically equivalent. As described previously, the actual vortex radius ($\xi$) and the site radius ($R_N$) are encoded together in the single parameter $r_c$  because this is the only relevant quantity (from the geometrical point of view) in the evaluation of the number of vortex--lattice interactions for a given configuration. However, the site radius in this picture has still a physical meaning: in order for a nucleus to \emph{actually} enter or exit the vortex core and thus change the vortex-lattice energy, the vortex axis must move by at least $2R_N$. Therefore,  if two extremal points are less than $ 2R_N$ apart there is no actual change in energy for the vortex to move from one to the other and therefore they must be counted together as a single pinning site. In other words we choose to count two extremal points as one if their distance is less than a quantity $\eta \sim 2 R_N$. 

 In conclusion, the number $\nNb(\tp) $ appearing in equation (\ref{eq:distance}) is corrected to take into account the ``clusters'' of extremal points as determined by the parameter $\eta$:  it corresponds to the number of  \emph{disjoint} clusters, each representing a physically distinct pinning site.
For the five zones in table \ref{tab:negele}, the quantity $2R_N$ is always in the range  $(0.25 \div 0.75)\Rws$. In order to make the calculations affordable, we fix $\eta = 0.5 \Rws$ for every zone, after testing that there is no significant difference in the final results for the pinning force  (below $10\%$ and anyways well within the error bars) under variations of $\eta$ in the previous range.  Altogether, it is evident that the main uncertainty in the calculation of the pinning force comes from the determination of $\nNb(\tp) $: in order to have some measure of this and since we are dealing with a counting measurement, we will associate to $\nNb$ the standard error $\pm\sqrt{\nNb}$.

To obtain the final value $ \f$ for the mesoscopic pinning force, we must repeat the above calculations for each value of $(\tp)$, and then take the angular average:
\begin{equation}
\label{eq:fl}
\f =\braket{\f(\tp)} =\frac{1}{4\pi}\int \f(\tp) \, d\Omega
\end{equation}
An estimate of the  error $\pm \s$ on $\f$ can also be obtained, by propagating  the error on $\nNb(\tp) $ in equations (\ref{eq:force})-(\ref{eq:fl}).

\begin{figure}
	\centering
	\includegraphics[width=8cm]{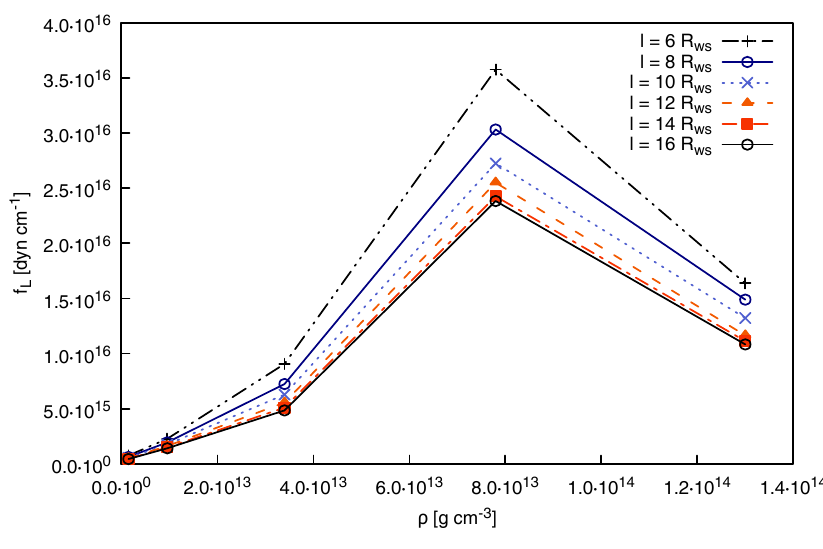}
	\caption{Convergence test for the $l$ parameter used in our calculation. In this figure we can see the pinning force per unit length for the five zones of the inner crust, obtained with different choices of the parameter: \emph{increasing} the value of $l$ the curves become closer, showing the convergence of the model. This picture corresponds to a vortex of length $L = 200 \Rws$.}
	\label{fig:conv_l}
\end{figure}

\begin{figure}
	\centering
	\includegraphics[width=8cm]{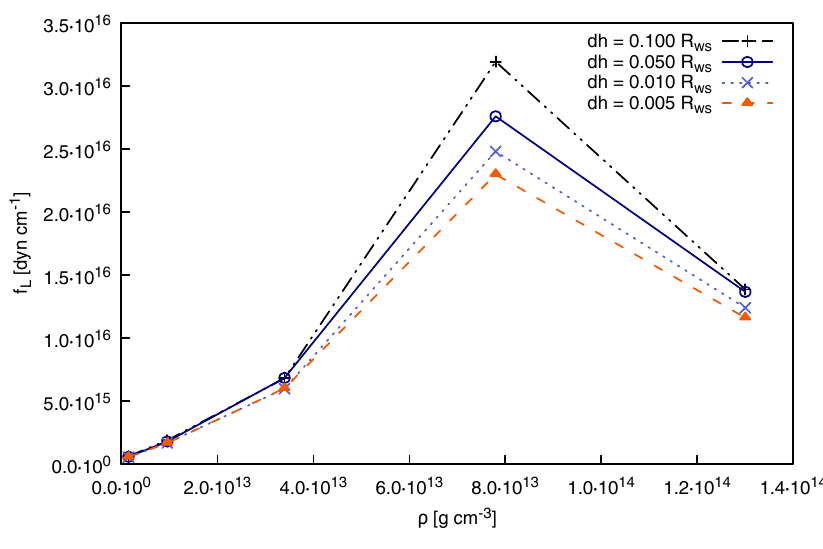}
	\caption{Convergence test for the $dh$ parameter used in our calculation. As in figure \ref{fig:conv_l} we can see that \emph{decreasing} the value of $dh$ the curves become closer, showing the convergence of the model also for this parameter.  This picture corresponds to a vortex of length $L = 200 \Rws$.}
	\label{fig:conv_dh}
\end{figure}

We also checked that our results are reasonably  independent from the choice of the parameters $l$ and $dh$ used in the parallel-transport sampling procedure.    In figures \ref{fig:conv_l} and \ref{fig:conv_dh} we show an example of  the convergence of  the calculated  $\f$ for different values of $l$ and $dh$. In the following we will fix $l=16 \,\Rws$ and $dh = 0.005 \, \Rws$, which provide an acceptable accuracy (well within the error bars $\pm \s$) while allowing for a not too long computational runtime.

\section{Results of the model: BCC lattice}

\begin{table*}
 \centering
 \caption{Results of the calculations for vortices with length $L$ up to $5000 \Rws$. The parameters of table \ref{tab:latticeprop} were used as inputs for the counting procedure. The quantities $\braket{D}$, $\braket{\Delta N}/\tilde{L}$ and $\braket{\Delta E}/L$ are the angular averages of $D(\tp)$,  $\Delta N(\tp)/\tilde{L}$ and $\Delta E(\tp)/L$ respectively. The last two columns show the force per unit length and its uncertainty.
}
 \label{tab:resZone} 
 \setlength{\tabcolsep}{14.5pt}
 \begin{tabular}{@{}rrrrrrrrr@{}}
 \hline
 $L$ & $\beta$ & $\#$ & $r_c$ & $\braket{D}$ & $\braket{\Delta N}/\tilde{L}$ & $\braket{\Delta E}/L$ & $\f$ & $\s$ \\
 $[R_{ws}]$ & & & $[R_{ws}]$ & $[R_{ws}]$ & [$10^{-2}$] & [$10^4$ erg/cm] & [$10^{15}$ dyn/cm] & [$10^{15}$ dyn/cm] \\
 \hline
  100 & 1 & 1 &    0.289 &    1.320 &    4.185 &    0.549 &    1.222 &    0.086 \\
      &   & 2 &    0.313 &    1.356 &    4.466 &    1.290 &    3.404 &    0.250 \\
      &   & 3 &    0.204 &    1.270 &   -3.687 &    5.689 &   22.970 &    1.453 \\
      &   & 4 &    0.531 &    2.064 &   -5.994 &   17.127 &   61.521 &    5.689 \\
      &   & 5 &    2.514 &    3.893 &  -13.779 &    5.439 &   12.578 &    2.501 \\
 \cline{2-9} \\
      & 3 & 1 &    0.591 &    2.072 &    6.730 &    0.417 &    0.630 &    0.069 \\
      &   & 2 &    0.555 &    2.005 &    6.496 &    0.704 &    1.362 &    0.143 \\
      &   & 3 &    0.504 &    2.060 &   -5.866 &    7.240 &   18.459 &    1.751 \\
      &   & 4 &    1.490 &    3.396 &   -9.725 &    5.542 &   10.904 &    1.698 \\
      &   & 5 &    8.080 &    4.634 &  -23.797 &    0.553 &    0.994 &    0.256 \\
 \cline{1-9} \\
  500 & 1 & 1 &    0.289 &    2.066 &    1.940 &    0.254 &    0.367 &    0.058 \\
      &   & 2 &    0.313 &    2.142 &    2.031 &    0.587 &    1.015 &    0.167 \\
      &   & 3 &    0.204 &    1.732 &   -1.504 &    2.321 &    7.190 &    0.833 \\
      &   & 4 &    0.531 &    2.880 &   -2.342 &    6.693 &   17.690 &    3.028 \\
      &   & 5 &    2.514 &    4.348 &   -5.184 &    2.046 &    4.349 &    1.277 \\
 \cline{2-9} \\
      & 3 & 1 &    0.591 &    3.092 &    2.900 &    0.179 &    0.191 &    0.046 \\
      &   & 2 &    0.555 &    2.970 &    2.809 &    0.304 &    0.416 &    0.095 \\
      &   & 3 &    0.504 &    2.804 &   -2.305 &    2.845 &    5.418 &    0.946 \\
      &   & 4 &    1.490 &    3.915 &   -3.541 &    2.018 &    3.529 &    0.726 \\
      &   & 5 &    8.080 &    4.842 &   -8.812 &    0.205 &    0.375 &    0.130 \\
 \cline{1-9} \\
 1000 & 1 & 1 &    0.289 &    2.407 &    1.431 &    0.188 &    0.238 &    0.053 \\
      &   & 2 &    0.313 &    2.467 &    1.491 &    0.431 &    0.651 &    0.152 \\
      &   & 3 &    0.204 &    2.049 &   -1.086 &    1.676 &    4.388 &    0.744 \\
      &   & 4 &    0.531 &    3.210 &   -1.615 &    4.616 &   11.033 &    2.587 \\
      &   & 5 &    2.514 &    4.467 &   -3.575 &    1.411 &    3.024 &    1.090 \\
 \cline{2-9} \\
      & 3 & 1 &    0.591 &    3.358 &    2.113 &    0.131 &    0.133 &    0.042 \\
      &   & 2 &    0.555 &    3.283 &    2.044 &    0.221 &    0.286 &    0.088 \\
      &   & 3 &    0.504 &    3.129 &   -1.606 &    1.982 &    3.403 &    0.816 \\
      &   & 4 &    1.490 &    4.131 &   -2.362 &    1.346 &    2.350 &    0.566 \\
      &   & 5 &    8.080 &    4.938 &   -6.037 &    0.140 &    0.267 &    0.112 \\
 \cline{1-9} \\
 2500 & 1 & 1 &    0.289 &    2.845 &    1.031 &    0.135 &    0.149 &    0.050 \\
      &   & 2 &    0.313 &    2.900 &    1.067 &    0.308 &    0.420 &    0.143 \\
      &   & 3 &    0.204 &    2.533 &   -0.756 &    1.167 &    2.462 &    0.688 \\
      &   & 4 &    0.531 &    3.544 &   -1.062 &    3.034 &    6.777 &    2.324 \\
      &   & 5 &    2.514 &    4.648 &   -2.355 &    0.930 &    2.184 &    0.981 \\
 \cline{2-9} \\
      & 3 & 1 &    0.591 &    3.663 &    1.513 &    0.094 &    0.096 &    0.040 \\
      &   & 2 &    0.555 &    3.530 &    1.455 &    0.158 &    0.207 &    0.083 \\
      &   & 3 &    0.504 &    3.468 &   -1.074 &    1.325 &    2.205 &    0.740 \\
      &   & 4 &    1.490 &    4.294 &   -1.463 &    0.833 &    1.506 &    0.466 \\
      &   & 5 &    8.080 &    4.992 &   -3.941 &    0.092 &    0.190 &    0.101 \\
 \cline{1-9} \\
 5000 & 1 & 1 &    0.289 &    3.067 &    0.852 &    0.112 &    0.123 &    0.049 \\
      &   & 2 &    0.313 &    3.147 &    0.884 &    0.255 &    0.339 &    0.140 \\
      &   & 3 &    0.204 &    2.783 &   -0.603 &    0.930 &    1.828 &    0.656 \\
      &   & 4 &    0.531 &    3.731 &   -0.828 &    2.366 &    5.317 &    2.233 \\
      &   & 5 &    2.514 &    4.681 &   -1.834 &    0.724 &    1.801 &    0.936 \\
 \cline{2-9} \\
      & 3 & 1 &    0.591 &    3.799 &    1.223 &    0.076 &    0.080 &    0.039 \\
      &   & 2 &    0.555 &    3.714 &    1.183 &    0.128 &    0.171 &    0.080 \\
      &   & 3 &    0.504 &    3.632 &   -0.844 &    1.042 &    1.677 &    0.713 \\
      &   & 4 &    1.490 &    4.482 &   -1.085 &    0.618 &    1.133 &    0.433 \\
      &   & 5 &    8.080 &    5.044 &   -2.996 &    0.070 &    0.153 &    0.096 \\
 \hline
 \end{tabular}
\end{table*}

\begin{figure*}
	\centering
	\includegraphics[width=14cm]{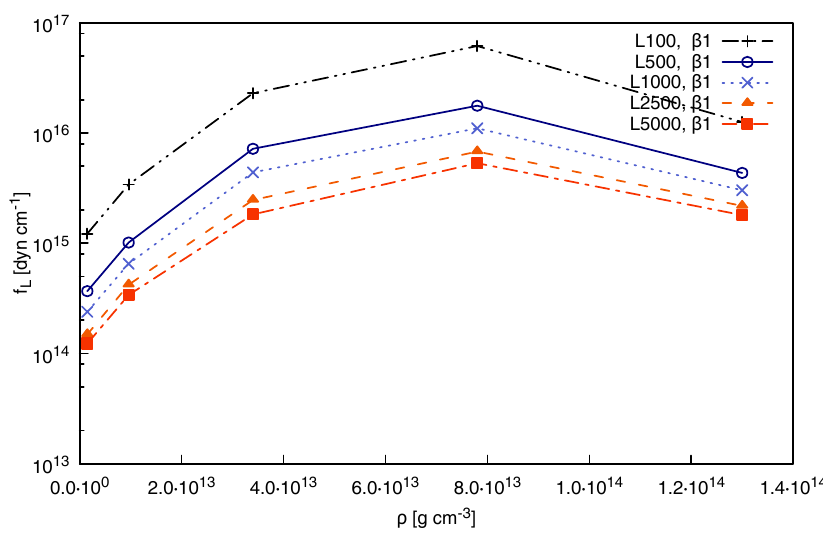}
	\caption{The pinning force per unit length for the $\beta = 1$ case. The mesoscopic pinning force is plotted as a function of the baryonic density of matter for the five zones considered and for different vortex lengths.} 
	\label{fig:curves_b1}
\end{figure*}

\begin{figure*}
	\centering
	\includegraphics[width=14cm]{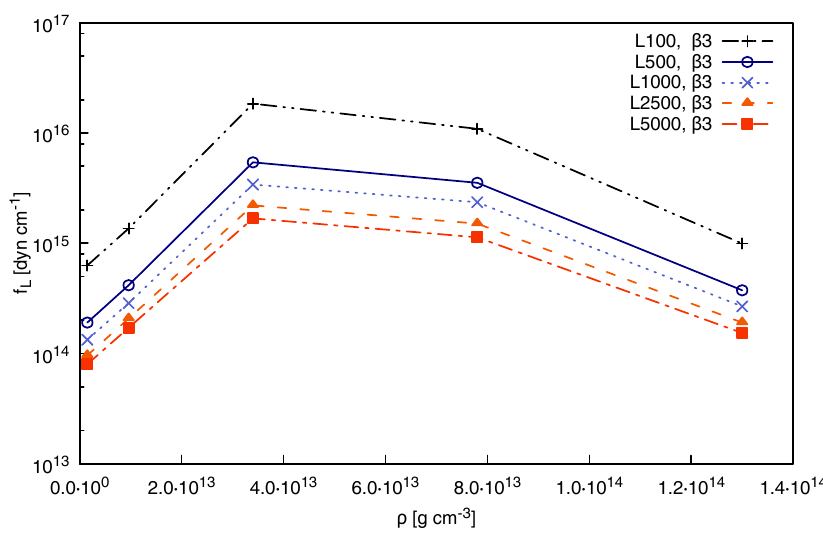}
	\caption{The pinning force per unit length for the $\beta = 3$ case. The pinning force per unit length for the $\beta = 1$ case. The mesoscopic pinning force is plotted as a function of the baryonic density of matter for the five zones considered and for different vortex lengths.}
	\label{fig:curves_b3}
\end{figure*}

The results of our calculations are summarized in table \ref{tab:resZone}. We have applied the algorithm described in the previous sections to different choices of the parameter $L$, starting from a short vortex with length equal to $100 \Rws$ up to a configuration with $L = 5000 \Rws$. For each value of $L$ and for each zone of table \ref{tab:negele} we have calculated the pinning force per unit length $\f$ and the estimated  error $\s$ for  two values of the polarization correction factor, $\beta=1$ (i.e. the case of a bare interaction) and $\beta=3$, which is close to the value obtained in  realistic Montecarlo simulations of neutron matter \citep{Gandolfi2008}.  The results for the pinning force per unit length are also plotted in figure \ref{fig:curves_b1} for  $\beta= 1$ and in figure \ref{fig:curves_b3} for $\beta=3$. 
 In the table we also show the results for $\braket{D}$, for $\braket{\Delta N}/\tilde{L}$ and for $\braket{\Delta E}/L$, which are the angular averages of $D(\tp)$,  $\Delta N(\tp)/\tilde{L}$, and $\Delta E(\tp)/L$ respectively.  We notice that $|\braket{\Delta N}|/\tilde{L} \ll 0.5$ when $\tilde{L} \sim 10^3$, which confirms the inadequacy of using symmetric vortex-lattice configurations when evaluating the mesoscopic pinning force \citep{Jones1990}.

From these results it is possible to see that there is a strong dependence of the pinning force per unit length on the parameter $L$: increasing the length of the vortex a consistent decrease in the mesoscopic pinning force can be observed. This behavior was indeed  expected, following the argument by \citet{jonesnopin} that the difference in energy between adjacent configurations becomes vanishingly small for infinite vortex rigidity ($L \rightarrow \infty$). However, using a realistic vortex length of order $\sim  10^3\Rws$, as discussed in section \ref{sec:vlen}, the pinning force is still not negligible.

The other important parameter of the model is the polarization factor $\beta$. The results show that $\f$ doesn't depend very strongly on the choice of this parameter in the three lower density regions, while the effect is more important in the two high density regions, where the mesoscopic pinning force is significantly larger  in the strong pairing scenario ($\beta = 1$) than in the weak one ($\beta = 3$).  It's also worth noting that changing the polarization factor from $\beta=1$ to $\beta=3$, results in a shift to lower densities of the maximum of the pinning profile. The position in density of the maximum pinning force can be relevant to determine the  angular momentum accumulated in the crust between pulsar glitches, as discussed in \citet{Pizzochero}.

\begin{figure*}
	\centering
	\includegraphics[width=14cm]{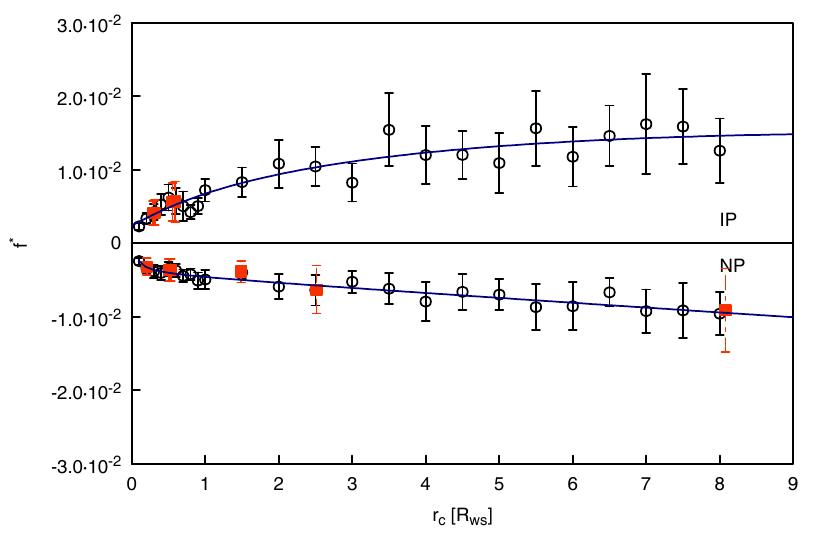}
	\caption{Plot of the calculated values of $\ff$ (for $\tilde{L} = 5000$) as a function of the capture radius $r_c$ (in units of $\Rws$). The error bars for the estimated errors on $\ff$ are also shown. The red squares are the values of $\ff$ corresponding to the ten values of $r_c$ in table \ref{tab:latticeprop}. The fitting curves $f^*$ for both the interstitial (above) and nuclear (below) pinning regimes   are also shown (see table \ref{tab:curves}).} 
	\label{fig:curves_5000}
\end{figure*}

A comparison between our results and those found in the literature shows that the maximum pinning forces per unit length obtained in this work are at least two orders of magnitude smaller than those found for an aligned vortex \citep{Alpar84a,Anderson82} and which have been commonly used in the study of pulsar glitches. Our results are also about one order of magnitude smaller than those obtained by \citet{link14} with a variational approach. Note that the main contribution to this difference derives from our estimate of the separation between pinned configurations, obtained from the counting procedure described above, which is larger than that used by \citet{link14}. As we shall discuss in the following, however, the forces we calculate are still large enough to account for the large glitches observed in the Vela pulsar.

Finally our calculations also provide an estimate of the fluctuations in local pinning strength that may be possible. The results in table \ref{tab:resZone} show that such fluctuations are generally of order $\approx 10\%$ of the pinning force, but can be larger in weaker pinning regions.

\subsection{Analytic approximations}

The results presented up to now refer to the calculation of the mesoscopic pinning force corresponding to the fiducial parameters $\Rws$, $R_N$, $\xi$ and $E_p$ in table  \ref{tab:negele}. However, existing or future calculations of the inner crust nuclear structure, of the neutron superfluid pairing properties and of the microscopic vortex-nucleus interaction may provide alternative sets of parameters to those used in the present work. It is possible to generalize our approach and  obtain a simple analytic expression which allows to calculate the pinning force per unit length $\f$ for different choices of the input parameters. 

In equation (\ref{eq:fl}), the quantity $\Eeff$ can be factorized. We can also express all the lengths in $\Rws$ units and then define an adimensional quantity $\ff(\tilde{r}_c)$ that depends \emph{only} on the adimensional capture radius $\tilde{r}_c = r_c/\Rws$ and the adimensional vortex length $\tilde{L}= L/\Rws$. The quantity $\ff(\tilde{r}_c)$ is purely geometrical and it contains all the information obtained from the counting procedure described in the preceding sections. The force per unit length $\f$ can then be obtained as 
\begin{equation}
\f=\ff\left(\tilde{r}_c\right) \frac{\Eeff}{\Rws^2}.
\label{eq:fstar}
\end{equation}

\begin{table}
	\centering
	\caption{Fit parameters for the function  $\model$. Three different vortex lengths $L$ are considered  for both the NP and IP regimes.}
	\label{tab:curves}
	\begin{tabular}{llrrrr}
	\hline
	 & & $A$ & $B$ & $W$ & $C$ \\
	 & & [$10^{-3}$] & [$10^{-3}$] & & [$10^{-3}$]\\
	 \hline                    
	 $L = 1000 \, \Rws$ & {NP} & -0.315 & -1.296  &  1.974  & -7.298 \\
		 				& {IP} & -2.099 &  9.043  &  1.586  &  7.212 \\
	 $L = 2500 \, \Rws$ & {NP} & -0.755 &  1.119  & -0.366  & -6.209 \\
	 					& {IP} & -0.374 &  7.685  &  0.997  &  3.057 \\
	 $L = 5000 \, \Rws$ & {NP} & -0.646 &  0.466  & -0.643  & -4.529 \\
	 					& {IP} & -0.772 &  7.641  &  1.114  &  2.428 \\
	\hline
	\end{tabular}
\end{table}

We have calculated $\ff(\tilde{r}_c)$ for different choices of $\tilde{r}_c$ (in the realistic range $0 \div 8$) and for different vortex lengths (of order $\tilde{L} \sim  10^3$) for both the NP and IP regimes
We then fitted a non-linear function $f^*(x)$ to the calculated values of $\ff$: we used a function of the form  
\begin{equation}
\label{eq:fit}
f^*(x) = Ax + B\left[ \log (1 + x) \right]^W + C
\end{equation}
where $A, B, C$ and $W$ are the parameters to be fitted.  In figure \ref{fig:curves_5000} we show the results for the $\tilde{L} = 5000$ case; the error bars have also been added, as obtained from the propagation of the error on $\nNb(\tp) $. We see that the calculated points can be fitted reasonably (within the error bars) with the choice of parameterization in equation (\ref{eq:fit}). In table \ref{tab:curves} we give the parameters obtained from the fitting procedure. 

We notice  that, within the uncertainty given by the quite large error bars, there is no significant difference in the magnitude of $\ff(\tilde{r}_c)$  between the {\em nuclear} and the {\em interstitial} regime. This means that the force per unit length, for  given $\tilde{r}_c$ and $\Eeff$, remains roughly the same if we  take the microscopic vortex-nucleus force to be attractive or repulsive. The fact that attractive and repulsive vortex-nucleus interactions are equivalent for the  pinning of  vortices to the lattice was already noted by \citet{link2009}.

\vspace{1cm}
\subsection{Vortex length}
\label{sec:length}
\begin{figure}
	\centering
	\includegraphics[width=8cm]{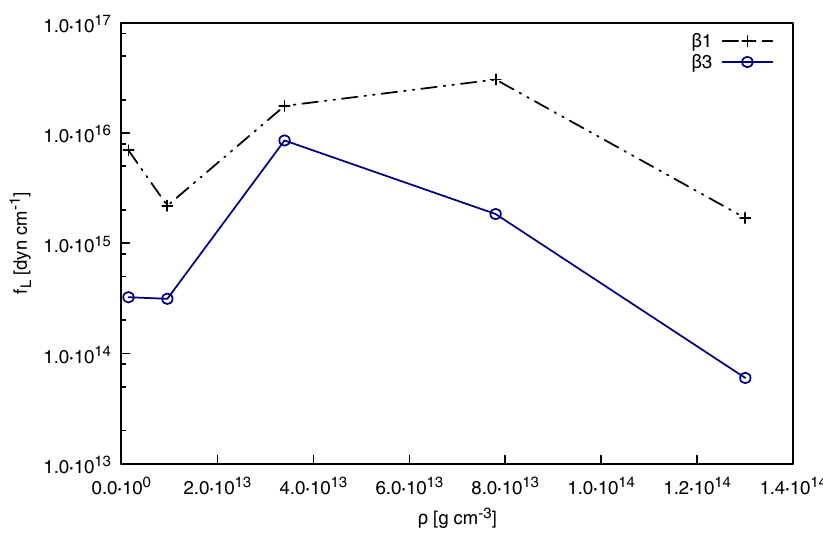}
	\caption{Plot of the values of pinning force per unit length reported in table \ref{tab:pinL}. These results are obtained by performing a fit for the dependence of $f_L$ on the rigidity length $L$ of the vortex (for each of the five zones considered), and then using equations (\ref{eq:length}) and (\ref{eq:tension}). }
	\label{fig:fitL}
\end{figure}

As discussed in section \ref{sec:vlen}, the parameter $L$  depends on the vortex tension, according to equation (\ref{eq:length}). The tension $T$ can be expressed as \citep{TF1, TF2, Andersson07}: 
\begin{equation}
\label{eq:tension}
T= \rho_n \frac{\kappa^2}{4 \pi} \log\left(\frac{a}{\xi}\right),
\end{equation}
where $\kappa = \pi \hbar / m_n$ is the quantum of circulation,  $a$ is the inter--vortex spacing and the neutron density is given by $\rho_n=x_n \rho$ with the neutron fraction $x_n\approx 0.9-0.95$ \citep{Zuo}. Given the logarithmic dependence on these parameters, we choose to expedite calculations and follow \citet{Andersson07} and take a constant value for the quantity $x_n \log\left(a/\xi\right) = 20 $ (note that a smaller value is considered by \citet{link14}). For each zone of the crust we calculate the tension and estimate the length $L$ over which we can consider the vortex as rigid. The dependence of the pinning force $f_L$ on the parameter $L$ is obtained by fitting, for every configuration ($\beta$, zone \#) considered in table \ref{tab:resZone}; a function of the form:
\begin{equation}
f = G \left[ \log \left(L/R_{ws}\right) \right]^H
\end{equation}
where $G$ and $H$ are the parameters to be fitted. In table \ref{tab:pinL} we report the results of the fit, together with the values of the vortex length $L$: we can see that at at lower densities the vortex is less rigid, with a significant enhancement in the rigidity taking place at high densities. The last column refers to the final result for the pinning force per unit length obtained by using the calculated value for $L$. These values are also plotted in figure \ref{fig:fitL}: the dashed line refers to the $\beta= 1$ condition, while the solid one is obtained with $\beta = 3$, which is a more realistic case, as indicated also by the results of \citet{Gandolfi2008}. Although the variable tension changes the profile of pinning with density  as compared to the case with constant tension, the mesoscopic pinning force still has values in the range $\f\approx 10^{14}-10^{15}$ dyn/cm for the realistic choice $\beta = 3$, while the largest densities (zone 5) correspond to superweak pinning \citep{Alpar84a}.

\begin{table}
	\centering
	\caption{Fitting results for the pinning force per unit length as a function of the vortex length $L$, for the five crustal zones (see text for details). The table reports also the best estimate for $L$ and the corresponding calculated value of $f_L$.}
	\label{tab:pinL}
\begin{tabular}{llrrrr}
	\hline
	$\beta$ & \# & $G$ & $H$ & $L$ & $f_L$ \\
	& & [$10^{19}\mbox{dyn/cm}$] & & $[R_{ws}]$ & [$10^{16}\mbox{dyn/cm}$]\\
	 \hline                    
	 $\beta = 1$ & 1 & 0.052 & -3.968 & 20    & 0.695 \\
	 			 & 2 & 0.151 & -3.990 & 171   & 0.220 \\
	   			 & 3 & 1.086 & -4.033 & 137   & 1.763 \\
	   			 & 4 & 3.527 & -4.159 & 232   & 3.055 \\
	 			 & 5 & 0.228 & -3.409 & 3984  & 0.169 \\
	 \hline
	 $\beta = 3$ & 1 & 0.019 & -3.740 & 244   & 0.032 \\
	 			 & 2 & 0.041 & -3.739 & 913   & 0.031 \\
	 			 & 3 & 0.925 & -4.071 & 260   & 0.855 \\
				 & 4 & 0.335 & -3.751 & 1634  & 0.184 \\
	 			 & 5 & 0.013 & -3.173 & 69726 & 0.006 \\
	\hline
	\end{tabular}
\end{table}

\section{Results of the model: random lattice}
\label{sec:random}

Given the uncertainties on the state of the crust at high densities, and the possibility that it may form a much less ordered structure than a BCC lattice \citep{Pethick}, let us consider the case of a random lattice, analogous to the case considered by \citet{link2009}.

The pinning force for a random lattice is calculated by applying the same procedure described for the BCC configuration. We consider a vortex of length $L$ inside a box of side $L + l$ where $l$ defines the area on which we perform the parallel transport operation, as done previously. As we intend to compare the results from the BCC configuration with this new setup, the box which represent our lattice must be filled with an adequate number $N_p$ of sites in order to obtain the same density. In a BCC lattice with  Wigner--Sietz cells of radius $R_{ws}$, the density of pinning sites is
$n_p = 1/(4 R_{ws}^3)$
and therefore the number of points we include in the random lattice, for comparison, must be
\be
                N_p = \frac{(L + l)^3}{4 R_{ws}^3}
\ee
The lattice is constructed by generating $N_p$ points extracted from a uniform distribution inside the box. For each orientation $(\theta, \phi)$ of the vortex we parallel transport it, and for each position we count the number of sites that fall inside the capture radius, as described previously, in order to evaluate the pinning force per unit length $F_L(\theta, \phi)$. Once we have generated a random lattice we keep it fixed for all orientations of the vortex.

\begin{figure}
	\centering
	\includegraphics[width=8cm]{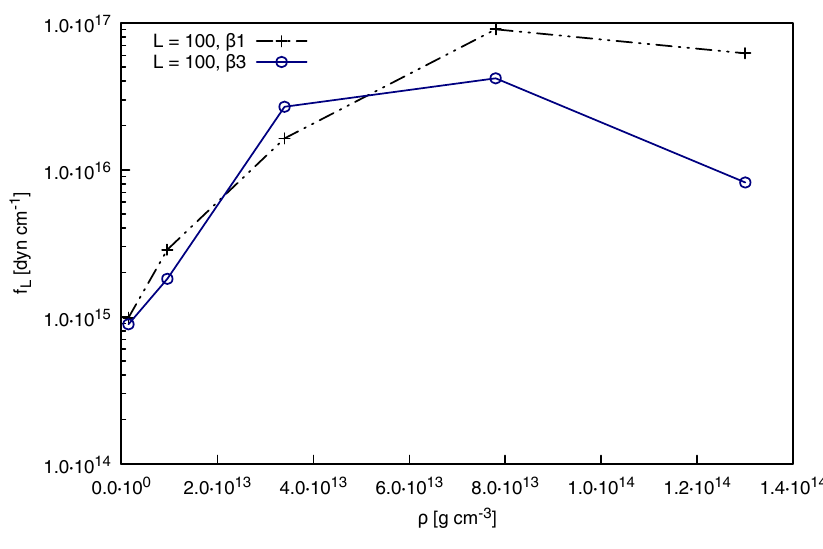}
	\caption{The pinning force per unit length for the $\beta = 1$ and $\beta=3$ case for a random lattice configuration. The mesoscopic pinning force is plotted as a function of the baryonic density of matter for the five zones considered. These results are referred to a vortex of length $L = 100 R_{ws}$. The order of magnitude of the pinning force is the same as in the BCC configuration.}
	\label{fig:random}
\end{figure}

The results for the different zones are shown in figure \ref{fig:random}. We can see that in general the order of magnitude of the pinning force is the same as in the BCC case, and appears to be determined by the average distance between pinning sites, the pinning energy, the coherence radius of the vortex and its rigidity at a given density, with the exact nature of the lattice only contributing a geometrical factor of order unity. Note that for a random lattice the distance between nuclei is, obviously, much more variable than for an ordered lattice, which will increase the error on our estimate of the bending lengthscale $L$. However, given the scaling of the pinning force as $f_L \sim 1/\sqrt{L}$,  upon averaging we still expect  an error  of less than a factor of a few in our estimates of the force itself.

The results presented here can also be explained by simple analytic considerations which involve Poisson statistics. For a random lattice configuration the average density of pinning sites is simply $n_N \approx 1/R_{ws}^3$ and therefore the number of sites inside a cylinder of radius $r_c$ and length $L$ is
\be
 N \approx \frac{L r_c^2}{R_{ws}^3}
 \ee
The numerical calculations presented above suggest that the typical distance $D$ between two maximal pinning configuration is of the order of $R_{ws}$, so we can estimate the force per unit length as $f_L = \Delta N \Eeff / (L R_{ws})$, with typical fluctuations in number of counted sites of $\Delta N = \sqrt{N}$ as we are considering a Poisson distribution. The final result is 
\begin{equation}
\label{eq:poisson}
f_L = \frac{1}{L} \frac{\Eeff}{R_{ws}}\frac{r_c}{R_{ws}}\sqrt{\frac{L}{R_{ws}}}   =\frac{\Eeff}{R_{ws}^2}\frac{\tilde{r}_c}{\sqrt{\tilde{L}}} = f^*\frac{\Eeff}{R_{ws}^2} 
\end{equation}
For $\tilde{r}_c \sim 1$ and $\tilde{L} \sim 10^3$, we obtain a reduction factor $f^* \sim 10^{-2}$ in agreement with the results of the previous sections.

Note also that the previous estimate for the pinning length has the same functional dependence on $L$, and thus on the tension, as the results of \citet{link}. We can thus interpret the average separation between pinning sites, $l_p$, introduced by \citet{linkcutler} and \citet{link} in our scheme. In our case $l_p$ represents the average distance between the additional `extra' pinning sites $\Delta N$ that lead to the difference between the bound and free energy configurations, namely  $\tilde{l}_p=l_p/R_{ws}$ is the inverse of the quantity $\braket{\Delta N}/\tilde{L}$ shown in table \ref{tab:resZone}. We remind the reader, however, that the vortex is never completely free from pinning sites and our 'free' configuration also contains pinning sites. The distance $l_p$ must thus be interpreted as a 'virtual' quantity between the $\Delta N$  excess pinning sites. As already discussed in section \ref{sec:vlen}, it is not the actual distance between physical nuclei, as in practice the average distance between nuclear pinning sites is always of order  $R_{ws}$.

\section{Application to pulsar glitches}

\begin{figure}
	\centering
	\includegraphics[width=8cm]{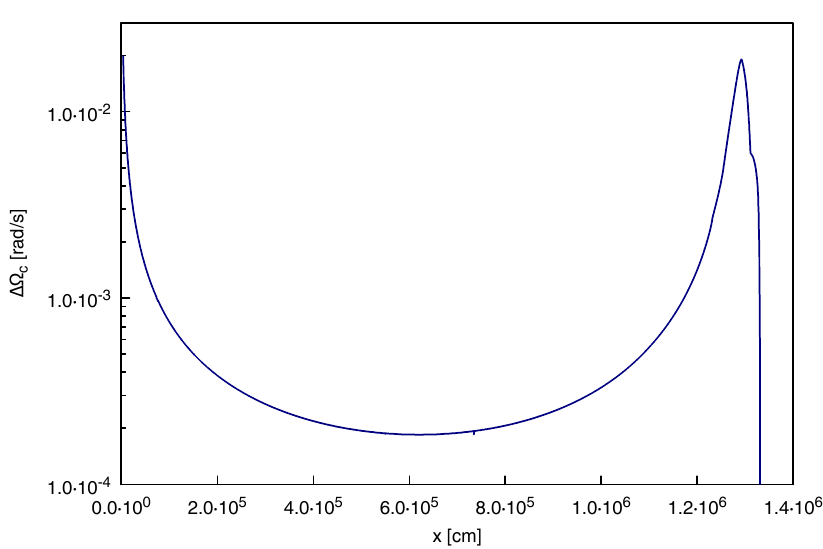}
	\caption{Critical unpinning lag $\Delta\Omega_c$ of a neutron star of $1.4M_\odot$. This profile has been obtained with a realistic model, by solving the general relativistic TOV equations and using the GM1 equation of state, as detailed in \citet{Seveso}.}
	\label{fig:crLag}
\end{figure}

Let us briefly outline how our results can impact on models of pulsar glitches. Let us consider a single pinned vortex: the forces acting on a section of it will be the pinning force calculated above $\mathbf{f}_L(\rho)$ and the Magnus force $\mathbf{f}_M=\kappa \rho_n \hat{\boldsymbol{\Omega}}\times (\mathbf{v}_n-\mathbf{v}_v)$, where $\hat{\boldsymbol{\Omega}}$ is a unit vector along the rotation axis, $\rho_n$ is the density of superfluid neutrons and $\mathbf{v}_n$ and $\mathbf{v}_v$ are the velocities of the superfluid neutrons and of the vortices respectively. Integrating these two contributions  over the full length of the vortex (which is taken to be straight), balancing  them and assuming that the pinned vortices move with the crust, allows us to determine, as a function of the distance from the rotational axis of the star, the \emph{critical lag} for unpinning $\Delta\Omega_{c}=\Omega_n-\Omega_p$, with $\Omega_p$ the angular velocity of the crust (the one which is observed). In figure \ref{fig:crLag} we show an example of the radial profile $\Delta\Omega_{c}$ for a typical $1.4 M_\odot$ NS, with the GM1 equation of state as detailed in \citet{Seveso}.
We can then follow the prescription of the 'snowplow' model of \citet{Pizzochero}, and assume that as a pulsar spins down vortices move out of the core and inner crust, and repin in the strong pinning region, eventually forming a vortex sheet close to the maximum $\Delta\Omega_{c_{\rm max}}$ of the critical lag.  
Given an equation of state and a critical  unpinning profile, we can therefore calculate the number of vortices involved in the process and the angular momentum stored by them, which eventually will power the glitch. It is also easy to evaluate the expected waiting time between glitches, that is the time needed to build the maximum critical lag: 
\begin{equation}
\label{eq:tgl}
t_{\rm gl} = \frac{\Delta\Omega_{{c}_{\rm max}}}{|\dot{\Omega}_p|},
\end{equation}
where $\dot{\Omega}_p$ is the observed pulsar spin down rate and for the moment we neglect effects of superfluid entrainment.

Following the approach of \citet{Pizzochero}, we can then compare the results of the present work to the giant glitches of the Vela pulsar:  we find that, for example, a typical observed glitch size of $\delta\Omega_p/\Omega_p\approx 10^{-6}$ can be obtained for a star of mass $1.3M_\odot$ with the GM1 equation of state, using the mesoscopic pinning  force corresponding to $L = 5000 \Rws$ and $ \beta = 3$ . It is also worth noting that  equation (\ref{eq:tgl}) gives for Vela a waiting time of $\sim 3$ years (in agreement with observational data), when the maximum of the mesoscopic pinning force is $f_{\rm{max}} \approx 10^{15} \mbox{ dyn/cm}$ (as it is the case for $L = 5000 \Rws$ and $ \beta = 3$ ). 
A more detailed study of the dependence of the snowplow model on parameters such as the equation of state and the mass of the star can be found in \citet{Seveso} and \citet{Haskell}. Effects of superfluid entrainment will also be considered in future work, as strong entrainment in the crust can severely limit the amount of angular momentum that is exchanged during a glitch and allow to set constraints on the equation of state  \citep{crust, chamelcrust}.

The simple model above shows that the pinning profiles we have calculated can play a very important role in the study of glitches, and could be used as a background for more realistic glitch models and vortex dynamics simulations \citep{Peralta06, Haskellglitch, Sideryglitch, Warszawski,Warszawski2, HaskellAnt}. Note that here we have only calculated the contribution to the pinning force acting on a vortex from the ions in the crust. In the core of the NS, however, protons are expected to form a type II superconductor, in which the magnetic field is organized in flux tubes, that can interact strongly and 'pin' the vortices \citep{Haskell}. We intend to apply the procedure described above to this scenario in a subsequent paper.

\section{Conclusions}

In this paper we present a calculation of the pinning force per unit length acting on a vortex in a neutron star crust. We have calculated the mesoscopic pinning force at different densities for straight vortices that cross the star inner crust, and averaged over all the possible orientations of the crustal lattice with respect to the vortex. Our results confirm the expectations of \citep{jonesnopin}, that the averaging procedure over different orientations tends to smooth out energy differences between different configurations, leading to weaker pinning forces. In the limit of infinitely long vortices the pinning force would vanish; for realistic values of the vortex tension and for the realistic choice $\beta=3$, however, the force per unit length is still sizable and in the range $\f\approx 10^{14}-10^{15}$ dyn/cm depending on the position in the crust and . These values are significantly smaller than those found in the literature; for instance, \citet{link14} obtains values typically of the order of $\f\approx 10^{16}$ dyn/cm. We find that the mesoscopic pinning force  depends very little on whether the pinning force is attractive (nuclear pinning) or repulsive (interstitial pinning) in a given region of the star, but it can be quite sensitive to in-medium polarization effects, which can shift the position of the maximum and thus alter the angular momentum distribution in the crust of a neutron star. Furthermore we have also considered the case of a more disordered crystal configuration than a BCC lattice and studied a random lattice.  In this case the pinning force does not vary significantly from the estimates in the BCC case, and depends mainly on the average distance between pinning sites, the pinning energy, the coherence length and rigidity of the vortex, with the exact nature of the lattice only contributing a geometric factor of order unity.

We also apply the calculated forces  to the problem of pulsar glitches and show how, in the framework of the 'snowplow' model \citep{Pizzochero}, our results can explain large glitches in the Vela pulsar. More generally the forces that we calculate can be used to generate realistic pinning profiles for glitch models \citep{Haskellglitch, HaskellAnt}, simulations of vortex dynamics in neutron stars \citep{Warszawski} or mode calculations \citep{hydro, link}. Future work will aim to include consistently the effect of strong crustal entrainment, as \citet{link14} has shown that including entrainment phenomenologically by rescaling the free neutron density can have important consequences for vortex creep, and more generally strong entrainment provides strong constraints for glitch models \citep{crust, chamelcrust}.

Finally let us note that we have considered the case of straight vortices that cross the star. Although this is the natural starting point for such a calculation, in a realistic neutron star the vortex array is likely to form a turbulent tangle (see e.g. \citet{Andersson07}) and pinning can also occur between vortices and superconducting flux tubes in the outer core of the neutron star, leading not only to an increased reservoir of angular momentum, but also to a modified response of the star to a glitch \citep{Haskell, SideryAlpar, Alinew}. 
Such configurations have a different topology from the one considered in this paper, and the behaviour of the pinning force in these cases is not captured by our current approach. In principle our calculation can, however, be modified to account for them.
Furthermore, as already mentioned, the crust of a neutron star may not form a BCC lattice but may exhibit a much more inhomogeneous structure \citep{Pethick} or exhibit several kinds of 'pasta' phases at the crust/core interface \citep{Ravenhall}, altering the geometry of the nuclear clusters. We intend to explore in detail the consequences of these effects on vortex pinning in future work.

Finally we note that our approach relies on calculating energy differences between vortex configurations and deriving from them a pinning force per unit length; therefore, it cannot give any information about the short-range radial profiles of the mesoscopic pinning energies and forces (incidentally, the existing  microscopic studies of the vortex-nucleus interaction are also based on energy arguments between specific configurations,  giving no information about the short-range radial dependence of the interaction). Dynamical vortex line simulations will thus be necessary to assess the stability of these configurations and how vortices move from one to another.

\section*{Acknowledgments}

BH acknowledges the support of the Australian Research Council (ARC) via a Discovery Early Career Researcher Award (DECRA) Fellowship. Partial support comes from ``NewCompStar'', COST Action MP1304.  SS thanks his PhD examination committee for interesting discussions and suggestions regarding pinning in a random lattice.
\nocite{*}
\bibliographystyle{mn2e}

\begin{thebibliography}{99}


\bibitem[\protect\citeauthoryear{Alpar}{1977}]{Alpar77} Alpar M.A., 1977, ApJ. 213, 527

\bibitem[\protect\citeauthoryear{Alpar et al.}{1984a}]{Alpar84a} Alpar M.A., Pines D., Anderson P.W., Shaham J., 1984a, ApJ. 276, 325

\bibitem[\protect\citeauthoryear{Alpar et al.}{1984b}]{Alpar84b} Alpar M.A., Anderson P.W., Pines D., Shaham J., 1984b, ApJ. 278, 791

\bibitem[\protect\citeauthoryear{Alpar}{1994}]{Alparquake} Alpar M.A., Chau, H.F., Cheng K.S., Pines D., 1994, ApJ. 427, L29

\bibitem[\protect\citeauthoryear{Alpar et al.}{1996}]{Alpardeplete} Alpar M.A., Chau, H.F., Cheng K.S., Pines D., 1996, ApJ. 459, 706

\bibitem[\protect\citeauthoryear{Anderson \& Itoh}{1975}]{Itoh} Anderson, P.W., Itoh N., 1975, Nature 256, 25

\bibitem[\protect\citeauthoryear{Anderson et al.}{1982}]{Anderson82} Anderson, P.W., Alpar M. A., Pines D., Shaham J., 1982, Phil. Mag. A, 45, 227

\bibitem[\protect\citeauthoryear{Andersson, Sidery \& Comer}{2007}]{Andersson07} Andersson N., Sidery T., Comer G.L., 2007, MNRAS 381, 747

\bibitem[\protect\citeauthoryear{Andersson et al.}{2012}]{crust} Andersson N., Glampedakis K., Ho W.C.G. Espinoza C.M., 2012, Phys. Rev. Lett. 109, 241103

\bibitem[\protect\citeauthoryear{Baym, Pethick, Pines \& Ruderman}{Baym et al.}{1969}]{Baym} Baym G., Pethick C., Pines D., Ruderman M., 1969, Nature 224, 872
\bibitem[\protect\citeauthoryear{Bildsten}{1998}]{Bild98} Bildsten L., 1998, ApJ. 501, L89

\bibitem[\protect\citeauthoryear{Chamel}{2012}]{chamelbragg} Chamel N., 2012, Phys. Rev. C 85, 035801
\bibitem[\protect\citeauthoryear{Chamel}{2013}]{chamelcrust} Chamel N., 2013, Phys. Rev. Lett 110, 011101

\bibitem[\protect\citeauthoryear{Donati \& Pizzochero}{Donati \& Pizzochero}{2003}]{DP03} Donati, P., Pizzochero P.M., 2003, Phys.Rev.Lett. 90, 21
\bibitem[\protect\citeauthoryear{Donati \& Pizzochero}{Donati \& Pizzochero}{2004}]{DP04} Donati, P., Pizzochero P.M., 2004, Nu.Phys.A, 742, 363
\bibitem[\protect\citeauthoryear{Donati \& Pizzochero}{Donati \& Pizzochero}{2006}]{DP06} Donati, P., Pizzochero P.M., 2006, Phys.Lett.B, 640

\bibitem[\protect\citeauthoryear{Elshamouty et al. }{2013}]{CASAnew} Elshamouty, K.G. et al., 2013, ApJ 777, 22
\bibitem[\protect\citeauthoryear{Epstein \& Baym}{Epstein \& Baym}{1988}]{EB} Epstein R.I.., Baym, G., 1988, ApJ. 328, 680

\bibitem[\protect\citeauthoryear{Espinoza et al. }{2011}]{Espinoza1} Espinoza, C.M., Lyne, A.G., Stappers, B.W., Kramer, M., 2011, MNRAS 414, 1679 

\bibitem[\protect\citeauthoryear{Fetter}{Fetter}{1967}]{TF2} Fetter, A.L., 1967, PhRvD 4, 1589 

\bibitem[\protect\citeauthoryear{Gandolfi et al.}{2008}]{Gandolfi2008} Gandolfi S., Illarionov A. Yu, Fantoni S., Pederiva F., Schmidt K. E., 2008, Phys. Rev. Lett., 101, 132501

\bibitem[\protect\citeauthoryear{Glampedakis \& Andersson}{Glampedakis \& Andersson}{2009}]{hydro} Glampedakis K., Andersson N., 2009, Phys. Rev. Lett. 102, 141101

\bibitem[\protect\citeauthoryear{Gudmundsson, Pethick \& Epstein}{1983}]{Gud} Gudmundsson E.H., Pethick C.J., Epstein R.I., 1983, ApJ 272, 286

\bibitem[\protect\citeauthoryear{G\"{u}gercino\u{g}lu \& Alpar}{2014}]{Alinew} G\"{u}gercino\u{g}lu, Alpar M.A., 2014, ApJ 788, L11

\bibitem[\protect\citeauthoryear{Haskell \& Melatos}{2015}]{HM15} Haskell B., Melatos A. 2015, submitted to MNRAS, arXiv:1510.03136

\bibitem[\protect\citeauthoryear{Haskell \& Antonopoulou}{2013}]{HaskellAnt} Haskell B., Antonopoulou D. 2014, MNRAS 438, L71

\bibitem[\protect\citeauthoryear{Haskell, Pizzochero \& Sidery}{2013}]{Haskellglitch} Haskell B., Pizzochero P.M., Sidery T. 2012, MNRAS 420, 658

\bibitem[\protect\citeauthoryear{Haskell, Pizzochero \& Seveso}{2013}]{Haskell} Haskell B., Pizzochero P.M., Seveso S., 2013 ApJ. 746 L25

\bibitem[\protect\citeauthoryear{Hirasawa \& Shibazaki}{2001}]{HS01} Hirasawa, M., Shibazaki,  N., 2001, ApJ 563, 267

\bibitem[\protect\citeauthoryear{Jones}{1990}]{Jones1990} Jones P. B., 1990, MNRAS 243, 257

\bibitem[\protect\citeauthoryear{Jones}{1991}]{jonesnopin} Jones P.B., 1991, ApJ. 373, 208

\bibitem[\protect\citeauthoryear{Kobyakov \& Pethick}{2014}]{Pethick} Kobyakov D., Pethick C., 2014, Phys. Rev. Lett. 112, 112504

\bibitem[\protect\citeauthoryear{Link \& Cutler}{2002}]{linkcutler} Link B., Cutler, C.,  2002, MNRAS 336, 211

\bibitem[\protect\citeauthoryear{Link}{2009}]{link2009} Link B., 2009, Phys. Rev. Lett. 102, 131101

\bibitem[\protect\citeauthoryear{Link}{2012}]{link} Link B., 2012, MNRAS 422, 1640

\bibitem[\protect\citeauthoryear{Link}{2014}]{link14} Link B., 2014, MNRAS 789, 141

\bibitem[\protect\citeauthoryear{Lorenz, Ravenhall \& Pethick}{1970}]{Ravenhall} Lorenz C.P., Ravenhall D.G., Pethick C.J., 1993, Phys. Rev. Lett. 70, 379

\bibitem[\protect\citeauthoryear{Middleditch et al.}{2006}]{quake2} Middleditch J., Marshall F.E., Wang Q.D., Gotthelf E.V., Zhang W., 2006, ApJ. 625, 1531

\bibitem[\protect\citeauthoryear{Migdal}{1959}]{Migdal} Migdal, 1959, ApJ. 743, L20

\bibitem[\protect\citeauthoryear{Negele \& Vautherin}{1973}]{Negele} Negele J.W, Vautherin D., 1973, Nucl. Phys. A. 207, 298

\bibitem[\protect\citeauthoryear{Page et al.}{2011}]{casA2} Page D., Parakash M., Lattimer J.M., Steiner, A.W., 2011, Phys. Rev. Lett. 106, 081101

\bibitem[\protect\citeauthoryear{Peralta}{2005}]{Peralta05} Peralta C., Melatos A., Giacobello M., Ooi, A., 2005, ApJ. 635, 1224

\bibitem[\protect\citeauthoryear{Peralta et al.}{2006}]{Peralta06} Peralta C., Melatos A., Giacobello M., Ooi, A., 2006, ApJ. 651, 1079


\bibitem[\protect\citeauthoryear{Piekarewickz, Fattoyev \& Horowitz}{2014}]{chuck} Piekarewicz J., Fattoyev F.J., Horowitz., 204 eprint: arXiv:1404.2660

\bibitem[\protect\citeauthoryear{Pizzochero}{2011}]{Pizzochero} Pizzochero P.M., 2011, ApJ. 743, L20



\bibitem[\protect\citeauthoryear{Ruderman}{1969}]{Rude} Ruderman M., 1969, Nature 223, 597

\bibitem[\protect\citeauthoryear{Ruderman}{1976}]{Rude2} Ruderman M., 1976, ApJ. 203, 213

\bibitem[\protect\citeauthoryear{Sedrakian}{1995}]{Sedrakian} Sedrakian A.D., 1995, MNRAS 277, 225 

\bibitem[\protect\citeauthoryear{Seveso et al.}{2012}]{Seveso} Seveso S.,  Pizzochero P.M., Haskell, B. 2012, MNRAS 427, 1089

\bibitem[\protect\citeauthoryear{Sidery, Passamonti \& Andersson}{2010}]{Sideryglitch} Sidery T., Passamonti A., Andersson N., 2010, MNRAS 405, 1061

\bibitem[\protect\citeauthoryear{Sidery \& Alpar}{2009}]{SideryAlpar} Sidery T., Alpar M.A., 2009, MNRAS 400, 1859

\bibitem[\protect\citeauthoryear{Shternin et al. }{2011}]{casA1} Shternin P.S., Yakovlev D. G., Heinke C.O., Ho W.C.G., Patnaude D.J., 2011, MNRAS 412, L108

\bibitem[\protect\citeauthoryear{Steiner et al.}{2014}]{steiner} Steiner A.W., Gandolfi S., Fattoyev F.J., Newton W.G., 2014, eprint:arXiv:1403.7546

\bibitem[\protect\citeauthoryear{Thomson}{Thomson}{1880}]{TF1} Thomson, W., 1880, PMAG 10, 155

\bibitem[\protect\citeauthoryear{Thompson \& Duncan}{Thompson \& Duncan}{1995}]{TD} Thompson C., Duncan R.C., 1995, MNRAS 275, 255

\bibitem[\protect\citeauthoryear{Warszawski \& Melatos}{2008}]{Warszawski} Warszawski L., Melatos A., 2008, MNRAS 390, 175

\bibitem[\protect\citeauthoryear{Warszawski \& Melatos}{2011}]{Warszawski2} Warszawski L., Melatos A., 2011, MNRAS 415, 1611

\bibitem[\protect\citeauthoryear{Warszawski \& Melatos}{2013}]{Warszawski3} Warszawski L., Melatos A., 2013, MNRAS 428, 1911

\bibitem[\protect\citeauthoryear{Zuo et al.}{2004}]{Zuo} Zuo W., Li Z.H., Lu G.C., Li J.Q., Scheid W., Lombardo U., Schulze H.-J., Shen C.W., 2004, Phys.Lett.B 595, 44


\end{thebibliography}

\appendix

\section{Vortex rigidity and unpinning}
\label{append}
\newcommand{\n}{\mathrm{n}}
\newcommand{\cc}{\mathrm{p}}
\newcommand{\LL}{\mathrm{L}}

Let us begin by writing the equations of motion for a single vortex. The forces acting on a vortex will be the Magnus force:
\be
f^i_M=\kappa \rho_\n \epsilon^{ijk} \hat{k}_j(v_k^\LL - v_k^\n),
\ee
and the non-dissipative part of the interaction with the pinning site, i.e. the `pinning' force ${F}^i_p$. For simplicity we neglect Mutual Friction in this example. The equations of motion for a vortex thus take the form:
\be
\epsilon^{ijk} \hat{k}_j(v_k^\LL - v_k^\n)+\mathcal{F}^i_p=0\label{gen}
\ee
where $v^\LL_i$ is the velocity of the vortex line and $v^\mathrm{n}_i$ is the velocity of the superfluid neutrons,  $\rho_\n$ the superfluid neutron mass density and $\mathcal{F}^i_p=F^i_p/{\rho_\n\kappa}$, with $\kappa$ the quantum of circulation (and $\hat{\kappa}_i$ the unit vector along the vorticity axis, taken to be the $z$ axis in the following).

As already discussed the large self energy of a vortex leads to tension due to flow around a curved segment, which introduces additional components in the individual neutron velocities, of the form \citep{Andersson07}:
\beq
\delta v_i^\n&=&\gamma_\n(\varepsilon)\nu\epsilon_{ijk}\hat{k}^j\hat{k}^p\nabla_p\hat{k}^k\label{ten1}
\eeq
where $\gamma$ is a function of the entrainment parameter $\varepsilon$ and 
\be
\nu=\frac{\kappa}{4\pi} \log{\left(\frac{a}{\xi}\right)}
\ee
with $a$ the inter vortex spacing and $\xi$ the coherence length associated with the vortex core, so that one has
\be
 \log{\left(\frac{a}{\xi}\right)}\approx 20 -\frac{1}{2} \log{\left(\frac{\Omega_\n}{100 \mbox{rad s$^{-1}$}}\right)}
 \ee
 which, as discussed in the main text, is essentially a constant for the range of periods of interest. Equation (\ref{eq:tension}) shows that the tension is $T = \nu\kappa\rho_n$.
It is thus sufficient to add the contributions in (\ref{ten1})  to the general flows in (\ref{gen}).

 For simplicity we follow \citet{Sedrakian} and  take a parabolic pinning potential $U_p$ of the form\footnote{As already noted, todate the short-range radial dependence of the vortex-nucleus interaction is not known.}:
\beq
U_p&=&\frac{1}{2}A(r_ir^i-r_0^ir_{0i})\;\;\;\mbox{for $|r_ir^i -r_0^ir_{0i}|\leq \Rrange$}\\
U_p&=&0\;\;\;\mbox{for $|r_ir^i -r_0^ir_{0i}|>\Rrange$}
\eeq
so that the pinning force $F_p^i=-\nabla^i U_p$ takes the form
\beq
F_p^i&=&-A(r^i-r_0^i)\;\;\;\mbox{for $|r_ir^i-r_0^ir_{0i}|\leq \Rrange$}\\
F_p^i&=&0\;\;\;\mbox{for $|r_ir^i-r_0^ir_{0i}|>\Rrange$}\label{force}
\eeq
where $A$ is a constant that describes the strength of the interaction and $r_0^i$ is the position of the pinning site, which for simplicity we shall take at the origin in the following, i.e. $r_0^i=0$. 

\begin{figure}
\centerline{\includegraphics[scale=0.4]{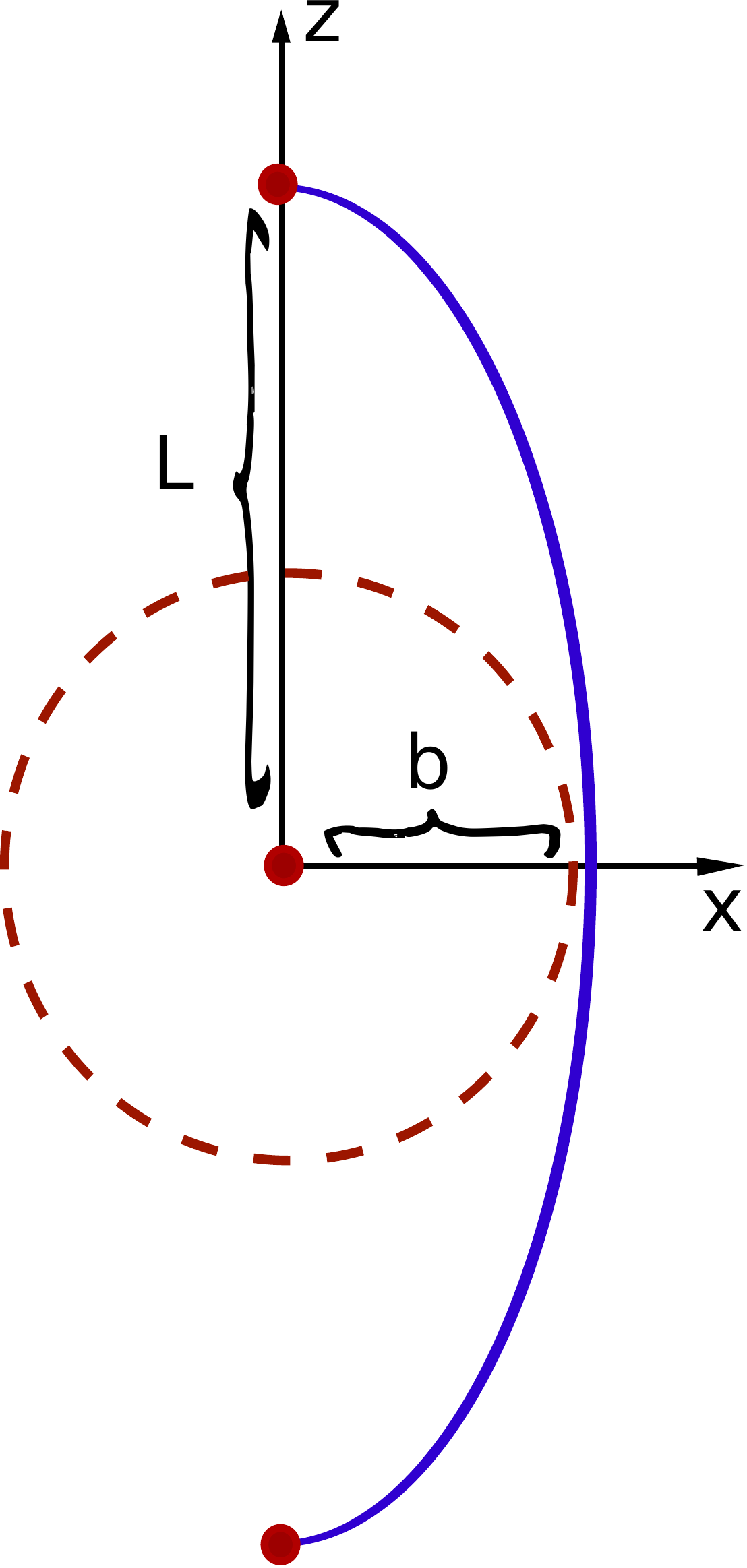}}
\caption{A schematic (out of scale) representation of a curved vortex, pinned at its extrema that has curved over length-scale $L$ to move away from a pinning site by a distance $b$. Red dots along the z axis indicate pinning sites at the unpinned edges of the vortex, and the dashed line shows the range of the pinning potential.}\label{ellipse}
\end{figure}

Consider a  pinned vortex that has to bend to free itself from one pinning bond. We work in a frame co-moving with the protons and take the background neutron velocity to be in the $y$ direction. To simplify our treatment we also assume that as a vortex moves out in the $x$ direction, driven by the Magnus force, it will take the shape of an ellipse in the x-z plane, as depicted in figure \ref{ellipse}, where $b$  is the distance from the centre of the pinning site on the x axis and L is the length  over which the vortex bends in the $z$ direction. Consider first a vortex that has unzipped from a single pinning site, so that $b\approx \Rrange$. The equations of motion for the vortex  take the form:
\be
v_i^\LL=V_i^\n+\epsilon_{ijk}\hat{k}^j\mathcal{F}^k-\hat{k}_i\hat{k}^j V_j^\n +\gamma_\n(\varepsilon)\nu\epsilon_{ijk}\hat{k}^j\hat{k}^p\nabla_p\hat{k}^k
\ee
where we have indicated as $V_i^\n$ the {\it background} superfluid neutron velocity (without the curvature induced contributions), and we take $\gamma_\n=1$ (which is appropriate in the crust for strong entrainment, see \citet{HM15} and \citet{chamelbragg}). Given that $\hat{k}_i$ has components only in the x-z plane it is sufficient to consider the vortex line velocity in the x direction:
\be
v_y^\LL=V_y^\n+(\hat{k}^z\mathcal{F}^x-\hat{k}^x\mathcal{F}^z)+\nu[\hat{k}^z\hat{k}_p\nabla^p\hat{k}_x-\hat{k}^x\hat{k}_p\nabla^p\hat{k}_z]\label{generalfree1}
\ee
which, evaluated at z=0 and at a point $x=b$ for the configuration in figure (\ref{ellipse}) leads to:
\be
v_y^\LL=V_y^\n-\mathcal{A}b-\nu\frac{b}{L^2}\label{tensionMf1}
\ee
where $\mathcal{A} = A / (\kappa\rho_n)$. Equation (\ref{tensionMf1}) shows that the tension acts in the same direction of the pinning force and tends to maintain the vortex straight. 
The critical velocity $\bar{V}_{cT}$ for unpinning with tension, compared to the critical velocity $V_{cr}$ in the absence of tension, is thus
\be
\bar{V}_{cT}=V_{cr}+\frac{\nu \Rrange}{L^2}\label{newpin}
\ee
where $\Rrange$ is the range of the pinning potential. 
Bending, and thus unpinning, over the length-scale of a single bond  is essentially prohibited by the tension,  since taking $L\approx \Rrange\approx \Rws$ leads to critical velocities of the order of $\bar{V}_{cT}\approx 10^8$ cm/s, far greater than what is achievable in a neutron star. From equation (\ref{newpin}) we see that, for a given pinning energy (i.e. fixed $V_{cr}$), the critical unpinning velocity has a minimum of $\bar{V}_{cT}=V_{cr}$ once $L$ is large enough. We can estimate that bending to unpin is possible when tension no longer increases the critical unpinning velocity. This will be the case if unpinning occurs over length-scales 
\be
L\gtrsim\sqrt{\frac{\nu \Rrange}{V_{cr}}}
\ee
and if we approximate the critical velocity as $V_{cr}\approx E_p/(\rho_\n\kappa \Rrange L)$ we find  
\be
L\gtrsim \frac{T \Rrange^2}{E_p}.
\ee
We can reasonably assume that $\Rrange \approx r_c $ and since $r_c \approx \Rws$ (within a factor of less than two for  $\beta=3$, see table \ref{sec:latticeprop}), we finally obtain  $L/ \Rws\approx (T \Rws)/E_p\approx 10^3$ for the unpinning length-scale in the deep crust, as estimated from energetics in section \ref{sec:vlen}.

\bsp
\label{lastpage}
\end{document}